\documentclass[12pt]{article}

\usepackage{amsmath}
\usepackage{amssymb}
\usepackage[dvips]{graphicx}
\usepackage{amsmath,amscd}
\usepackage{color}
\usepackage{mathrsfs}
\usepackage{latexsym}
\usepackage{bm}
\usepackage{fourier}
\usepackage{multirow}
\allowdisplaybreaks[4]
\makeatletter

  \@addtoreset{equation}{section}
\makeatother
\allowdisplaybreaks[4]

\begin{document}
\begin{titlepage}
\thispagestyle{empty}
\begin{flushright}

\end{flushright}
\vspace{3mm}

\begin{center}
{\LARGE Orbifold family unification 
using vectorlike representation on six dimensions}

\end{center}

\begin{center}
\lineskip .45em
\vskip1.5cm
{\large Yuhei Goto$^a$\footnote{E-mail: y-goto@keio.jp}
and Yoshiharu Kawamura$^b$
\footnote{E-mail: haru@azusa.shinshu-u.ac.jp}

\vskip 1.5em
${}^a\,${\large\itshape Research and Education Center 
for Natural Science, Keio University, }\\
{Yokohama 223-8522, Japan,}\\[1mm]
${}^b\,${\large\itshape Department of Physics, Shinshu University, }\\
{Matsumoto 390-8621, Japan}
}


\vskip 4.5em
\end{center}
\begin{abstract}
In orbifold family unification 
on the basis of $SU(N)$ gauge theory 
on the six-dimensional space-time
$M^4\times T^2/Z_m$ ($m=2, 3, 4, 6$),
enormous numbers of models 
with three families of the standard model matter multiplets
are derived from a massless Dirac fermion 
in a vectorlike representation $[N, 3] + [N, N-3]$ of $SU(N)$
($N = 8, 9$).
They contain models with three or more than three neutrino singlets
and without any non-Abelian continuous flavor gauge symmetries.
The relationship
between flavor numbers
from a fermion in $[N, N-k]$ and those from a fermion in $[N, k]$
are studied from the viewpoint of charge conjugation.
\end{abstract}

\end{titlepage}
\newpage

\abovedisplayskip=1.0em
\belowdisplayskip=1.0em
\abovedisplayshortskip=0.5em
\belowdisplayshortskip=0.5em

\section{Introduction}

One of the most intriguing riddles in particle physics
is the origin of the family replication
in the standard model (SM) matter multiplets.
Various investigations have been performed,
using models on the four-dimensional Minkowski 
space-time $M^4$~\cite{R,ss3,G,F1,F2,W&Z,K&Y},
but, in most cases,
we encounter difficulties 
relating to the chiralness of fermions.
Concretely, chiral fermions do not, in general, come from
a fermion in an anomaly free representation of
a large gauge group, e.g., ${\bf 2}^{n-1}$ for $SO(2n)$ ($n \ge 6$),
or a vectorlike (non-chiral) set of representations, e.g., 
${\bm N}+ \overline{\bm N}$ for $SU(N)$,
as an extension of grand unified theories (GUTs).
In most cases, particles 
with opposite quantum numbers 
under the SM gauge group $SU(3)_C \times SU(2)_L \times U(1)_Y$,
called mirror particles, appear
and the survival hypothesis is adopted
to get rid of them from the low-energy spectrum.
Then, the SM family members can also disappear.
Here, the survival hypothesis is stated such that 
{\it if a symmetry is broken down into a smaller one 
at a scale $M_S$,
then any fermion mass terms invariant 
under the smaller group induce fermion masses 
of $O(M_S)$
and such heavy fermions 
disappear from the low-energy spectrum}~\cite{G,BNM&S}.
\footnote{
There is a possibility that extra particles are confined 
at a high-energy scale by some strong dynamics~\cite{ss3,W&Z}.
}

The above difficulty can be overcome
by extending the structure of space-time.
That is, extra particles including mirror ones 
can be eliminated using orbifold breaking mechanism,
as originally proposed 
in superstring theory~\cite{CHS&W,DHV&W1,DHV&W2}.
Hence, a candidate realizing the family unification
is an extension of GUTs
defined on a higher-dimensional space-time
including an orbifold.\footnote{
Five-dimensional supersymmetric GUTs 
on $M^4 \times S^1/Z_2$ possess the attractive feature that
the triplet-doublet splitting of Higgs multiplets 
is elegantly realized~\cite{K,H&N}.}
These studies have been carried out 
intensively~\cite{BB&K,Watari:2002tf,
GMN1,GLMN,GLMS,GMN2,KK&O,K&M,FKMN&S,GK&M,Yamatsu1,Yamatsu2},
and three replicas of matter multiplets
are derived from characteristics of extra dimensions.
For instance, 
three replication $SU(5)$ multiplets 
have been derived from a single bulk 
fermion in the rank $k$ totally antisymmetric tensor representation
$[N, k]$ ($N \ge 9$) of $SU(N)$
on $M^4 \times S^1/Z_2$~\cite{KK&O}.
Enormous numbers of models 
with three families of the SM matter multiplets
have been obtained from a single massless Dirac fermion 
in $[N, k]$ ($N \ge 9$) of $SU(N)$ on $M^4 \times T^2/Z_m$ 
($m=2, 3, 4$)~\cite{GK&M}.
The relationship between the flavor numbers of chiral fermions
and the Wilson line phases has been studied
in these models~\cite{GK&M2}.
Using models originated from $SU(9)$ gauge theory
on $M^4 \times T^2/Z_2$,
their reality has been examined
from the structure of the Yukawa interactions~\cite{G&K}.

In Ref.~\cite{GK&M},
we find that the number of neutrino singlets
is less than three, the smallest gauge group is $SU(9)$,
and most models contain extra non-Abelian 
continuous gauge group relating to 
a flavor symmetry, under the precondition that three SM families
are derived from a massless Dirac fermion in 
a chiral representation $[N, k]$ of $SU(N)$.
Then, we need extra neutrino singlets to produce 
massive neutrinos and extra scalar fields
to break extra gauge symmetries.
By changing the precondition into 
that three SM families
are derived from a massless Dirac fermion
in a vectorlike representation $[N, k] + [N, N-k]$ of $SU(N)$,
there is a possibility that some models possess features 
such that
the number of neutrino singlets
is three or more than three, 
the smallest gauge group is less than $SU(9)$,
and all extra gauge symmetries are Abelian.
Furthermore, extra gauge symmetries could be broken down 
by the vacuum expectation values of 
superpartners of neutrino singlets.

In this paper, we study the possibility of family unification 
on the basis of $SU(8)$ and $SU(9)$ gauge theory 
on $M^4\times T^2/Z_m$, 
using the method in Ref.~\cite{KK&O,GK&M}.
We investigate whether or not three families of 
the SM matter multiplets
are derived from a single massless Dirac fermion 
in a vectorlike representation $[8, k] + [8, 8-k]$
or $[9, k] + [9, 9-k]$, 
through the orbifold breaking mechanism.
We clarify the relationship
between flavor numbers
from a fermion in $[N, N-k]$ 
and those from a fermion in $[N, k]$
from the viewpoint of charge conjugation.

The contents of this paper are as follows.
In Sec. 2, we provide general arguments on the orbifold breaking 
based on two-dimensional orbifold $T^2/Z_m$.
In Sec. 3, we give formulae for numbers of the SM matter multiplets.
In Sec. 4, we study a possibility of the family unification
in six-dimensional $SU(8)$ and $SU(9)$ gauge theories
containing a massless Dirac fermion in a vectorlike representation. 
Section 5 is devoted to conclusions and discussions.

\section{$Z_m$ orbifold breaking, fermions
and decomposition of field}
\label{S2}

We explain the orbifold $T^2/Z_m$ ($m=2, 3, 4, 6$),
a six-dimensional fermion
and a decomposition of field in $[N, k]$.

\subsection{$Z_m$ orbifold breaking}
\label{S2-1}

On a two-dimensional lattice $T^2$, 
the points $z + e_1$ and $z + e_2$ are identified with
the point $z$, where $e_1$ and $e_2$ are basis vectors
and $z$ takes a complex value.
The orbifold $T^2/Z_m$ is obtained 
by dividing $T^2$ by the $Z_m$ transformation
$z \to \rho z$, where $\rho$ is the $m$-th root of unity $(\rho^m = 1)$.
Then, $z$ is identified with $\rho z$, 
or $z$ is identified with
$\rho^k z + a e_1 + b e_2$, where $k$, $a$ and $b$ are integers.
For more details, see Appendix A.

We explain the $Z_m$ transformation properties
of a six-dimensional scalar field $\Phi(x, z, \overline{z})$,
using $T^2/Z_3$
whose basis vectors are given by $e_1 = 1$ and $e_2 = i$.
The extension of other fields (fermions and gauge bosons)
and other orbifolds is straightforward.
From the requirement that the Lagrangian density $\mathscr{L}$
should be invariant under the $Z_3$ transformations 
$s_0 : z \to \omega z$ 
and $s_1 : z \to \omega z + 1$ ($\omega = e^{2\pi i/3}$)
or it should be a single-valued function,
\begin{eqnarray}
\mathscr{L}(\Phi(x, \omega z, \overline{\omega}~\overline{z}))
= \mathscr{L}(\Phi(x, z, \overline{z})),~~ 
\mathscr{L}(\Phi(x, \omega z + 1, \overline{\omega}~\overline{z} + 1))
= \mathscr{L}(\Phi(x, z, \overline{z})),
\label{L-single}
\end{eqnarray}
the boundary conditions of fields on $T^2/Z_3$
are determined up to some overall $Z_3$ factors,
which we refer to as intrinsic $Z_3$ elements of fields
and denote as $\eta_{a\Phi}$
corresponding to the $Z_3$ transformations $s_a$ ($a=0, 1$).
When $\Phi$ is a multiplet of some transformation group $G$
concerning some internal symmetries (including gauge symmetries),
$\mathscr{L}$ should be invariant under the transformation
$\Phi(x, z, \overline{z}) \to 
\Phi'(x, z, \overline{z}) = T_{\Phi} \Phi(x, z, \overline{z})$,
such that
\begin{eqnarray}
\mathscr{L}(T_{\Phi}\Phi(x, z, \overline{z}))
=\mathscr{L}(\Phi(x, z, \overline{z})),
\label{L-inv}
\end{eqnarray}
where $T_{\Phi}$ is a representation matrix of $G$ on $\Phi$.
For instance, if a theory has $SU(N)$ gauge symmetry,
$\mathscr{L}$ is, in general, invariant under a (global) $U(N)$ transformation,
i.,e., $G = U(N)$.
From (\ref{L-single}) and (\ref{L-inv}),
the following boundary conditions on $\Phi$ are allowed,
\begin{eqnarray}
\Phi(x, \omega z, \overline{\omega}~\overline{z}) 
=T_{\Phi}[U_0, \eta_{0\Phi}] \Phi(x, z, \overline{z}),~~ 
\Phi(x, \omega z + 1, \overline{\omega}~\overline{z} + 1) 
=T_{\Phi}[U_1, \eta_{1\Phi}] \Phi(x, z, \overline{z}),
\label{T[U]}
\end{eqnarray}
where $T_{\Phi}[U_0, \eta_{0\Phi}]$ and $T_{\Phi}[U_1, \eta_{1\Phi}]$ 
represent appropriate representation matrices,
which are elements of $G$ on $\Phi$.
The $T_{\Phi}[U_a, \eta_{a\Phi}]$ are factorized into
$T_{\Phi}[U_a, \eta_{a\Phi}]=\eta_{a\Phi} \tilde{T}_{\Phi}[U_a]$ ($a = 0, 1$),
using representation matrices $U_a$ 
for the fundamental representations of $G$
and the intrinsic $Z_3$ elements $\eta_{a\Phi}$ (see (\ref{etaNka})),
and some relations can appear among the intrinsic $Z_3$ elements
(see (\ref{rhopsi}) or (\ref{eta-psi})).
Arbitrary $U_0$ and $U_1$ can be diagonalized simultaneously
by a global unitary transformation
and a local gauge transformation or each equivalence class
of boundary conditions contains diagonal representatives~\cite{Yamashita}.
Hence we use diagonal ones later.

We list basis vectors and the transformations relating to
identifications of points on $T^2/Z_m$,
and denote its representation matrices 
for the fundamental representation as $U_a$ 
($a=0, 1, 2$ for $T^2/Z_2$ and $a=0, 1$ for $T^2/Z_3$ and
$T^2/Z_4$ and $a=0$ for $T^2/Z_6$),
in Table \ref{character}~\cite{K&M3,K&M2}.
\begin{table}[htb]
\caption{The characters of $T^2/Z_m$.}
\label{character}
\begin{center}
\begin{tabular}{c|c|c|c} \hline
$T^2/Z_m$ & Basis vectors & Transformations 
& Representation matrices \\ \hline\hline
$T^2/Z_2$ & $1, i$  
& $z \to -z,~ z \to 1-z,~ z \to i-z$
& $U_0,~ U_1,~ U_2$ \\ \hline
$T^2/Z_3$ & $1, e^{2\pi i/3}$  
& $z \to e^{2\pi i/3} z,~ z \to e^{2\pi i/3} z + 1$ 
& $U_0,~ U_1$ \\ \hline
$T^2/Z_4$ & $1, i$ & $z \to iz,~ z \to iz + 1$ & $U_0,~ U_1$ \\ \hline
$T^2/Z_6$ & $1, (-3+i\sqrt{3})/2$  
& $z \to e^{\pi i/3} z$ & $U_0$ \\ \hline
\end{tabular}
\end{center}
\end{table}
Note that there is a choice in transformations
independently of each other.

Components of $\Phi$ possess discrete charges 
associated with eigenvalues of $T_{\Phi}[U_a, \eta_{a\Phi}]$.
When the eigenvalues are given as $e^{2\pi i l/m}$
($l=0, 1, \cdots, m-1$),
the discrete charges are assigned as numbers $l/m$.
We refer to $e^{2\pi i l/m}$ as $Z_m$ elements.
In the absence of contributions from the Wilson line phases,
the massless six-dimensional fields whose $Z_m$ elements 
for all $a$ are
equal to $1$ contain zero modes, 
but those including a $Z_m$ element
different from $1$ do not contain zero modes.\footnote{
In the presence of non-vanishing Wilson line phases,
gauge symmetries and particle spectrum are rearranged
via the Hosotani mechanism~\cite{Hosotani1,Hosotani2,HHH&K,HH&K}.
}
Here, zero modes mean four-dimensional massless fields. 
If the size of extra dimensions is small enough,
massive modes called Kaluza-Klein modes 
do not appear in low-energy theories.
{\it Unless all components of non-singlet field have 
a common $Z_m$ charge, 
a symmetry reduction occurs upon compactification.}
This type of symmetry breaking mechanism is called 
$\lq\lq$orbifold breaking mechanism''.\footnote{
The $Z_2$ orbifolding was used in superstring theory~\cite{A} 
and heterotic M-theory~\cite{H&W1,H&W2}.
In field theoretical models, it was applied to 
the reduction of global supersymmetry (SUSY)~\cite{M&P,P&Q}, which is
an orbifold version of Scherk-Schwarz mechanism~\cite{S&S,S&S2}, 
and then to the reduction of gauge symmetry~\cite{K1}.
}

\subsection{Fermions}
\label{S2-2}

We explain fermions in six dimensions.
For more details, see the Appendix B.
A massless Weyl fermion on six dimensions 
is regarded as a Dirac fermion 
or a pair of Weyl fermions with opposite chiralities on four dimensions.
The six-dimensional Dirac fermion consists of 
two six-dimensional Weyl fermions such that
\begin{eqnarray}
\Psi_+ = \frac{1+\Gamma_7}{2} \Psi 
= \left( 
\begin{array}{c}
\psi_{+L} \\
\psi_{+R} 
\end{array}
\right),~~
\Psi_- = \frac{1-\Gamma_7}{2} \Psi
= \left( 
\begin{array}{c}
\psi_{-R} \\
\psi_{-L} 
\end{array}
\right),
\label{Psi}
\end{eqnarray}
where $\Psi_+$ and $\Psi_-$ are fermions with positive
and negative chirality, respectively, 
and $\Gamma_7$ is the chirality operator on six dimensions.
Here and hereafter, the subscript $\pm$ and $L(R)$
stand for the chiralities on six and four dimensions, respectively.
The charge conjugation of 
a six-dimensional Dirac fermion $\Psi$ is defined as
\begin{eqnarray}
\Psi^c \equiv B \Psi^*,~~ B^{-1} \Gamma^{M} B 
= -\left(\Gamma^M\right)^*,
\label{CC}
\end{eqnarray}
where $\Gamma^M$ ($M=0, 1, 2, 3, 5, 6$)
are six-dimensional gamma matrices,
$B = -i\Gamma_7 \Gamma^2 \Gamma^5$
up to a phase factor, and the asterisk $*$ means the complex
conjugation. \footnote{
In this paper, the complex conjugation is also represented
by the overlined one.
}
Note that the chirality in six dimensions
does not flip under the charge conjugation,
as shown in (\ref{Psi+CC}) and (\ref{Psi-CC}). 

From the $Z_m$ invariance of kinetic term
and the transformation property of the covariant derivatives 
$D_z \to \overline{\rho} D_z$ 
and $D_{\overline{z}} \to \rho D_{\overline{z}}$
with $\overline{\rho}(={\rho}^*) = e^{-2\pi i/m}$
 and $\rho = e^{2\pi i/m}$,
we have the relations:
\begin{eqnarray}
\eta_{a+R} = {\rho} \eta_{a+L},~~
\eta_{a-R} = \overline{\rho} \eta_{a-L},
\label{rhopsi}
\end{eqnarray}
where $z \equiv x^5 + i x^6$ and $\overline{z} \equiv x^5 - i x^6$,
and $\eta_{a\pm L(R)}$ are 
the intrinsic $Z_m$ elements of $\psi_{\pm L(R)}$.
For the derivation of (\ref{rhopsi}),
see from (\ref{Psi+kinetic}) to (\ref{eta-psi}).

Chiral gauge theories including Weyl fermions 
on even dimensional space-time
become, in general, anomalous in the presence of gauge anomalies,
gravitational anomalies, mixed anomalies 
and/or global anomaly~\cite{D&P,BG&T}.
Here we consider a non-supersymmetric model for simplicity.
In $SU(N)$ gauge theories on six dimensions,
the global anomaly is absent 
because of $\pi_6(SU(N)) = 0$ for $N \ge 4$.
Here, $\pi_6(SU(N))$ is the six-th homotopy group of $SU(N)$.
Other anomalies must be canceled out 
by the contributions from several fermions.
For instance, they are canceled out by the contributions from fermions 
with different chiralities such as $(\Psi^{\bm{r}}_+, \Psi^{\bm{r}}_-)$,
where $\bm{r}$ stands for $r$-dimensional representation
of $SU(N)$. 
Each pair in 
$(\Psi^{\bm{r}}_+, \Psi^{\overline{\bm{r}}}_-)$, 
$(\Psi^{\overline{\bm{r}}}_+, \Psi^{\bm{r}}_-)$
and $(\Psi^{\overline{\bm{r}}}_+, \Psi^{\overline{\bm{r}}}_-)$
does not contribute to the anomalies,
where $\overline{\bm{r}}$ stands for 
the complex conjugate representation of $\bm{r}$.
The cancellation on six dimensions
is understood that 
the gauge anomaly is proportional to a group-theoretical factor
such as
\begin{eqnarray}
\sum_{\Psi_+} \mbox{Str}\left(T^{a_1} T^{a_2} T^{a_3} T^{a_4}\right)
- \sum_{\Psi_-} \mbox{Str}\left(T^{a_1} T^{a_2} T^{a_3} T^{a_4}\right),
\label{Str}
\end{eqnarray}
where Str stands for the trace over the symmetrized product
of the gauge group generators $T^{a_i}$, and
this trace is invariant under the exchange between 
$T^{a_i}$ and $-(T^{a_i})^*$, corresponding to
the exchange between
a fermion in $\bm{r}$ and one in $\overline{\bm{r}}$.
The gravitational anomaly is canceled out,
if the following condition is fulfilled,
\begin{eqnarray}
N_{+} = N_{-},
\label{gr-cond}
\end{eqnarray}
where $N_{\pm}$ is the numbers (including degrees of freedom)
of $\Psi_{\pm}$.

\subsection{Decomposition of representation}
\label{S2-3}

With suitable diagonal representation matrices $U_a$,
the $SU(N)$ gauge group is broken down into its subgroup such that
\begin{eqnarray}
SU(N) \to SU(p_1)\times SU(p_2) \times \cdots \times 
SU(p_{n})\times U(1)^{n-n'-1},
\label{GSB}
\end{eqnarray}
where $N=p_1 + p_2 + \cdots + p_n$.
Here and hereafter, $SU(1)$ unconventionally stands for 
$U(1)$, $SU(0)$ means nothing
and $n'$ is a sum of the number of $SU(0)$. 
A concrete form of 
$U_a$ will be given in the next section.

After the breakdown of $SU(N)$, 
the rank $k$ totally antisymmetric tensor representation $[N, k]$, 
whose dimension is ${}_{N}C_{k}$,
is decomposed into a sum of multiplets of the subgroup 
$SU(p_1) \times SU(p_2) \times \cdots \times SU(p_n)$ as
\begin{eqnarray}
[N, k] = \sum_{l_1 =0}^{k} \sum_{l_2 = 0}^{k-l_1} 
\cdots \sum_{l_{n-1} = 0}^{k-l_1-\cdots -l_{n-2}}  
\left({}_{p_1}C_{l_1}, {}_{p_2}C_{l_2}, \cdots, {}_{p_n}C_{l_n}\right),
\label{Nk}
\end{eqnarray}
where $l_n=k-l_1- \cdots -l_{n-1}$ 
and our notation is that ${}_{n}C_{l} = 0$ for $l > n$ and $l < 0$.
Here and hereafter, we use ${}_{n}C_{l}$ instead of $[n, l]$ in many cases.
We sometimes use the ordinary notation for representations too, 
e.g., $\bm{N}$ and $\overline{\bm{N}}$ 
in place of ${}_{N}C_{1}$ and ${}_{N}C_{N-1}$. 

The $[N, k]$ is constructed 
by the antisymmetrization of $k$-ple product of 
the fundamental representation $\bm{N} = [N, 1]$:
\begin{eqnarray}
[N, k] = (\underbrace{\bm{N} \times \dots 
\times \bm{N}}_k)_{\mbox{\tiny{A}}},
\label{N*...*N} 
\end{eqnarray}
where a tiny subscript A means
the antisymmetrization.
For Weyl fermions $\Psi_{\pm}$ 
in $[N, k]$, the boundary conditions are given by
\begin{eqnarray}
\Psi_{\pm}(x, \rho z, \overline{\rho}~\overline{z}) 
=T_{\Psi\pm}[U_a, \eta^{(k)}_{a\pm}] \Psi_{\pm}(x, z, \overline{z}),
\label{TPsi}
\end{eqnarray}
where $T_{\Psi\pm}[U_a, \eta^{(k)}_{a\pm}]$ stand for appropriate
representation matrices, which are elements of $U(N)$ on $\Psi_{\pm}$,
$U_a$ are the representation matrices
for the fundamental representation
and $\eta^{(k)}_{a\pm}$ are the intrinsic $Z_m$ elements
of $\Psi_{\pm}$ in $[N, k]$.
We omit the subscripts $L$ and $R$ on $\eta^{(k)}_{a\pm}$,
for simplicity.
Note that there are relations such as (\ref{rhopsi})
between $\eta^{(k)}_{a\pm L}$ and $\eta^{(k)}_{a\pm R}$.
Using (\ref{N*...*N}) and (\ref{TPsi}),
the $Z_m$ transformation property of $[N, k]$ can be expressed by
\begin{eqnarray}
(\bm{N} \times \dots \times \bm{N})_{\mbox{\tiny{A}}}
~\to~
\eta^{(k)}_{a\pm} ((U_a {\bm{N}}) \times \dots 
\times (U_a {\bm{N}}))_{\mbox{\tiny{A}}}.
\label{etaNka}
\end{eqnarray}
By definition, $\eta^{(k)}_{a\pm}$ take values of $Z_m$ elements, 
i.e., $e^{2\pi i l/m}$ ($l=0, 1, \cdots, m-1$).
Note that $\eta^{(k)}_{a+}$ are not necessarily 
same as $\eta^{(k)}_{a-}$,
and the chiral symmetry is still respected.

In the same way, the $[N, N-k]$ is constructed 
by the antisymmetrization of $(N-k)$-ple product of $\bm{N}$:
\begin{eqnarray}
[N, N-k] = (\underbrace{\bm{N} \times \dots 
\times \bm{N}}_{N-k})_{\mbox{\tiny{A}}},
\label{N*...*N-(N-k)} 
\end{eqnarray}
or it is also constructed 
by the antisymmetrization of $k$-ple product of 
the complex conjugate representation $\overline{\bm{N}}$:
\begin{eqnarray}
[N, N-k] = \overline{[N, k]}
= (\underbrace{\overline{\bm{N}} \times \dots 
\times \overline{\bm{N}}}_k)_{\mbox{\tiny{A}}}.
\label{N*...*N-*} 
\end{eqnarray}
Using (\ref{N*...*N-*}),
the $Z_m$ transformation property is given by
\begin{eqnarray}
(\overline{\bm{N}} \times \dots 
\times \overline{\bm{N}})_{\mbox{\tiny{A}}}
~\to~
\tilde{\eta}^{(k)}_{a\pm} 
((U^*_a \overline{\bm{N}}) \times \dots 
\times (U^*_a \overline{\bm{N}}))_{\mbox{\tiny{A}}},
\label{etaNka-*}
\end{eqnarray}
where $U^*_a$ are the complex conjugations of $U_a$, and
$\tilde{\eta}^{(k)}_{a\pm}$ are the intrinsic $Z_m$ elements
of $\Psi_{\pm}$ in $\overline{[N, k]}$.
If the field in $\overline{[N, k]}$ is obtained
by the charge conjugation of that in $[N, k]$,
we have relations 
$\tilde{\eta}^{(k)}_{a\pm} = \overline{\eta^{(k)}_{a\pm}}$.
Strictly speaking, in this case, 
the relations are written as 
$\tilde{\eta}^{(k)}_{a\pm R} = \overline{\eta^{(k)}_{a\pm L}}$
and $\tilde{\eta}^{(k)}_{a\pm L} = \overline{\eta^{(k)}_{a\pm R}}$,
because the four-dimensional chirality
changes under the charge conjugation.
If a field in $[N, k]$ is independent of 
that in $\overline{[N, k]}$,
there is no relation between 
$\eta^{(k)}_{a\pm}$ and $\tilde{\eta}^{(k)}_{a\pm}$.

\section{Formulae for numbers of SM species}
\label{S3}

Let us investigate the family unification with
the breaking pattern:
\begin{eqnarray}
SU(N) \to  SU(3) \times SU(2) \times SU(p_3) \times 
\cdots \times SU(p_n) \times U(1)^{n-n'-1}~,
\label{OB}
\end{eqnarray}
where $SU(3)$ and $SU(2)$ are identified 
with $SU(3)_C$ and $SU(2)_L$ in the SM gauge group.
After the breakdown of $SU(N)$, $[N, k]$ is decomposed 
into a sum of multiplets as
\begin{eqnarray}
[N, k] = \sum_{l_1 =0}^{k} \sum_{l_2 = 0}^{k-l_1} \sum_{l_3 = 0}^{k-l_1-l_2} 
\cdots \sum_{l_{n-1} = 0}^{k-l_1-\cdots -l_{n-2}}  
\left({}_{3}C_{l_1}, {}_{2}C_{l_2}, {}_{p_3}C_{l_3}, \cdots, {}_{p_n}C_{l_n}\right).
\label{Nk(p1=3)}
\end{eqnarray}

The flavor numbers of down-type anti-quark singlets $(d_{R})^c$, 
lepton doublets $l_{L}$, up-type anti-quark singlets $(u_{R})^c$, 
positron-type lepton singlets $(e_{R})^c$, 
and quark doublets $q_{L}$ are denoted as 
$n_{\bar{d}}$, $n_l$, $n_{\bar{u}}$, $n_{\bar{e}}$ and $n_q$.
Using the survival hypothesis 
and the equivalence on charge conjugation in four dimensions, 
we define the flavor number of each SM chiral fermion as
\begin{eqnarray}
&~& n_{\bar{d}} \equiv 
\left(\sharp ({}_{3}C_{2}, {}_{2}C_{2})_L 
- \sharp  ({}_{3}C_{1}, {}_{2}C_{0})_L\right) 
- \left(\sharp  ({}_{3}C_{2}, {}_{2}C_{2})_R
- \sharp  ({}_{3}C_{1}, {}_{2}C_{0})_R\right),  
\label{nd-def}\\
&~& n_{l} \equiv 
\left(\sharp  ({}_{3}C_{3}, {}_{2}C_{1})_L  
- \sharp  ({}_{3}C_{0}, {}_{2}C_{1})_L\right)
- \left(\sharp  ({}_{3}C_{3}, {}_{2}C_{1})_R 
- \sharp  ({}_{3}C_{0}, {}_{2}C_{1})_R\right),  
\label{nl-def}\\
&~& n_{\bar{u}} \equiv 
\left(\sharp  ({}_{3}C_{2}, {}_{2}C_{0})_L  
- \sharp  ({}_{3}C_{1}, {}_{2}C_{2})_L\right)
- \left(\sharp  ({}_{3}C_{2}, {}_{2}C_{0})_R
- \sharp  ({}_{3}C_{1}, {}_{2}C_{2})_R\right),  
\label{nu-def}\\
&~& n_{\bar{e}} \equiv 
\left(\sharp  ({}_{3}C_{0}, {}_{2}C_{2})_L  
- \sharp  ({}_{3}C_{3}, {}_{2}C_{0})_L\right) 
- \left(\sharp  ({}_{3}C_{0}, {}_{2}C_{2})_R 
- \sharp  ({}_{3}C_{3}, {}_{2}C_{0})_R\right),  
\label{ne-def}\\
&~& n_{q} \equiv 
\left(\sharp  ({}_{3}C_{1}, {}_{2}C_{1})_L  
- \sharp  ({}_{3}C_{2}, {}_{2}C_{1})_L\right) 
- \left(\sharp  ({}_{3}C_{1}, {}_{2}C_{1})_R
- \sharp  ({}_{3}C_{2}, {}_{2}C_{1})_R\right),  
\label{nq-def}
\end{eqnarray}
where $\sharp$ represents the number of zero modes for each multiplet.
The SM singlets are regarded as the right-handed neutrinos, 
which can obtain heavy Majorana masses among themselves 
as well as the Dirac masses with left-handed neutrinos.
Some of them can be involved in see-saw mechanism~\cite{ss1,ss2,ss3}.
The total number of (heavy) neutrino singlets $(\nu_{R})^c$ 
and/or $\nu_{R}$
is denoted by $n_{\bar{\nu}}$ and defined as
\begin{eqnarray}
n_{\bar{\nu}} \equiv \sharp  ({}_{3}C_{0}, {}_{2}C_{0})_L  
+ \sharp  ({}_{3}C_{3}, {}_{2}C_{2})_L + \sharp  ({}_{3}C_{0}, {}_{2}C_{0})_R 
+ \sharp  ({}_{3}C_{3}, {}_{2}C_{2})_R.  
\label{nnu-def}
\end{eqnarray}

From (\ref{Nk(p1=3)}), the number of zero modes for each multiplet
is given by the formulae:
\begin{eqnarray}
\sharp ({}_{3}C_{l_1}, {}_{2}C_{l_2})_L &=&  
\sum_{l_3 = 0}^{k-l_1-l_2} 
\cdots \sum_{l_{n-1} = 0}^{k-l_1-\cdots -l_{n-2}} P_{mk\pm L}~
{}_{p_3}C_{l_3} \cdots {}_{p_{n}}C_{l_{n}}, 
\label{k1k2-L}\\
\sharp ({}_{3}C_{l_1}, {}_{2}C_{l_2})_R &=&  
\sum_{l_3 = 0}^{k-l_1-l_2} 
\cdots \sum_{l_{n-1} = 0}^{k-l_1-\cdots -l_{n-2}} P_{mk\pm R}~
{}_{p_3}C_{l_3} \cdots {}_{p_{n}}C_{l_{n}},
\label{k1k2-R}
\end{eqnarray}
where the $P_{mk\pm L(R)}$ ($m=2, 3, 4, 6$) are projection operators
to pick out zero modes of $\psi_{\pm L(R)}$ in $[N, k]$,
and they are listed in Table \ref{projection}.
\begingroup
\renewcommand{\arraystretch}{1.2}
\begin{table}[htb]
\caption{The projection operators $P_{mk\pm L(R)}$.}
\label{projection}
\begin{center}
\begin{tabular}{c|c|c|c|c} \hline
$T^2/Z_m$ & $P_{mk+L}$ & $P_{mk+R}$ & $P_{mk-L}$ & $P_{mk-R}$
\\ \hline\hline
$T^2/Z_2$ & $P_{2k+}^{(1,1,1)}$ & $P_{2k+}^{(-1,-1,-1)}$
& $P_{2k-}^{(1,1,1)}$ & $P_{2k-}^{(-1,-1,-1)}$
\\ \hline
$T^2/Z_3$ & $P_{3k+}^{(1,1)}$ 
& $P_{3k+}^{(\overline{\omega}, \overline{\omega})}$
& $P_{3k-}^{(1,1)}$ & $P_{3k-}^{(\omega, \omega)}$
\\ \hline
$T^2/Z_4$ & $P_{4k+}^{(1,1)}$ 
& $P_{4k+}^{(-i, -i)}$
& $P_{4k-}^{(1,1)}$ & $P_{4k-}^{(i, i)}$
\\ \hline
$T^2/Z_6$ & $P_{6k+}^{(1)}$ 
& $P_{6k+}^{(\overline{\varphi})}$
& $P_{6k-}^{(1)}$ & $P_{6k-}^{(\varphi)}$
\\ \hline
\end{tabular}
\end{center}
\end{table}
\endgroup
In Table \ref{projection}, $\varphi = e^{i\pi/3}$
and $\overline{\varphi} = e^{-i\pi/3}$,
and each operator is defined by
\begin{eqnarray}
&~& P_{2k\pm}^{\left((-1)^{n_0}, (-1)^{n_1}, (-1)^{n_2}\right)} \equiv 
\frac{1}{8}\left\{1 + (-1)^{n_0} \mathcal{P}^{(k)}_{0 \pm}\right\}
\left\{1 + (-1)^{n_1} \mathcal{P}^{(k)}_{1 \pm}\right\}
\left\{1 + (-1)^{n_2} \mathcal{P}^{(k)}_{2 \pm}\right\},
\label{P2}\\
&~& P_{3k\pm}^{\left(\omega^{n_0}, \omega^{n_1}\right)} 
\equiv \frac{1}{9}
\left\{1+ \overline{\omega}^{n_0} \mathcal{P}^{(k)}_{0 \pm} 
+ \overline{\omega}^{2n_0} \left(\mathcal{P}^{(k)}_{0 \pm}\right)^2\right\}
\left\{1+ \overline{\omega}^{n_1} \mathcal{P}^{(k)}_{1 \pm} 
+ \overline{\omega}^{2n_1} \left(\mathcal{P}^{(k)}_{1 \pm}\right)^2\right\},
\label{P3}\\
&~& P_{4k\pm}^{\left(i^{n_0}, i^{n_1}\right)} 
\equiv \frac{1}{16}
\left\{1+ (-i)^{n_0} \mathcal{P}^{(k)}_{0 \pm} 
+ (-i)^{2n_0} \left(\mathcal{P}^{(k)}_{0 \pm}\right)^2
+ (-i)^{3n_0} \left(\mathcal{P}^{(k)}_{0 \pm}\right)^3\right\}
\nonumber \\
&~& ~~~~~~~~~~~~~~~~~~~~~~~~~
\times \left\{1+ (-i)^{n_1} \mathcal{P}^{(k)}_{1 \pm} 
+ (-i)^{2n_1} \left(\mathcal{P}^{(k)}_{1 \pm}\right)^2
+ (-i)^{3n_1} \left(\mathcal{P}^{(k)}_{1 \pm}\right)^3\right\},
\label{P4}\\
&~& P_{6k\pm}^{\left(\varphi^{n_0}\right)} 
\equiv \frac{1}{6}
\left\{1+ \overline{\varphi}^{n_0} \mathcal{P}^{(k)}_{0 \pm} 
+ \overline{\varphi}^{2n_0} \left(\mathcal{P}^{(k)}_{0 \pm}\right)^2
+ \overline{\varphi}^{3n_0} \left(\mathcal{P}^{(k)}_{0 \pm}\right)^3
+ \overline{\varphi}^{4n_0} \left(\mathcal{P}^{(k)}_{0 \pm}\right)^4
+ \overline{\varphi}^{5n_0} \left(\mathcal{P}^{(k)}_{0 \pm}\right)^5
\right\},
\label{P6}
\end{eqnarray}
where $n_0$, $n_1$ and $n_2$ are integers,
$\mathcal{P}^{(k)}_{a \pm}$ are 
the $Z_m$ elements determined by
$U_a$ and $\eta^{(k)}_{a \pm L(R)}$, as will be given below.
For instance, $P_{3k\pm}^{\left(\omega^{n_0}, \omega^{n_1}\right)}$
is an projection operator to pick out
modes with $\mathcal{P}^{(k)}_{0 \pm} = \omega^{n_0}$
and $\mathcal{P}^{(k)}_{1 \pm} = \omega^{n_1}$ in $\Psi_{\pm}$.

From (\ref{nd-def}) -- (\ref{nnu-def}), (\ref{k1k2-L}) and (\ref{k1k2-R}),
we obtain following formulae for the SM species
and neutrino singlets 
derived from a pair of six-dimensional Weyl fermions $(\Psi_{+}, \Psi_{-})$
in $[N, k]$,
\begin{eqnarray}
\left. n_{\bar{d}}\right|_{[N, k]} &=& \sum_{\pm} \sum_{(l_1, l_2) = (2,2),(1,0)} 
\sum_{l_3 = 0}^{k-l_1-l_2} 
\cdots \sum_{l_{n-1} = 0}^{k-l_1-\cdots -l_{n-2}} 
(-1)^{l_1+l_2} P_{mk\pm}~
{}_{p_3}C_{l_3} \cdots {}_{p_{n}}C_{l_{n}}, 
\label{nd-ZM}\\
\left. n_{l}\right|_{[N, k]} &=& \sum_{\pm} \sum_{(l_1, l_2) = (3,1),(0,1)} 
\sum_{l_3 = 0}^{k-l_1-l_2} 
\cdots \sum_{l_{n-1} = 0}^{k-l_1-\cdots -l_{n-2}} 
(-1)^{l_1+l_2}  P_{mk\pm}~
{}_{p_3}C_{l_3} \cdots {}_{p_{n}}C_{l_{n}}, 
\label{nl-ZM}\\
\left. n_{\bar{u}}\right|_{[N, k]} &=& \sum_{\pm} \sum_{(l_1, l_2) = (2,0),(1,2)} 
\sum_{l_3 = 0}^{k-l_1-l_2} 
\cdots \sum_{l_{n-1} = 0}^{k-l_1-\cdots -l_{n-2}} 
(-1)^{l_1+l_2} P_{mk\pm}~
{}_{p_3}C_{l_3} \cdots {}_{p_{n}}C_{l_{n}}, 
\label{nu-ZM}\\
\left. n_{\bar{e}}\right|_{[N, k]} &=& \sum_{\pm} \sum_{(l_1, l_2) = (0,2),(3,0)} 
\sum_{l_3 = 0}^{k-l_1-l_2} 
\cdots \sum_{l_{n-1} = 0}^{k-l_1-\cdots -l_{n-2}} 
(-1)^{l_1+l_2} P_{mk\pm}~
{}_{p_3}C_{l_3} \cdots {}_{p_{n}}C_{l_{n}}, 
\label{ne-ZM}\\
\left. n_{q}\right|_{[N, k]} &=& \sum_{\pm} \sum_{(l_1, l_2) = (1,1),(2,1)} 
\sum_{l_3 = 0}^{k-l_1-l_2} 
\cdots \sum_{l_{n-1} = 0}^{k-l_1-\cdots -l_{n-2}} 
(-1)^{l_1+l_2} P_{mk\pm}~
{}_{p_3}C_{l_3} \cdots {}_{p_{n}}C_{l_{n}}, 
\label{nq-ZM}\\
\left. n_{\bar{\nu}}\right|_{[N, k]} &=& \sum_{\pm} \sum_{(l_1, l_2) = (0,0),(3,2)}
\sum_{l_3 = 0}^{k-l_1-l_2} 
\cdots \sum_{l_{n-1} = 0}^{k-l_1-\cdots -l_{n-2}} 
P_{mk\pm}^{(\nu)}~{}_{p_3}C_{l_3} \cdots {}_{p_{n}}C_{l_{n}},
\label{nnu-ZM}
\end{eqnarray}
where $P_{mk\pm}$ and $P_{mk\pm}^{(\nu)}$ are defined by
\begin{eqnarray}
P_{mk\pm} \equiv P_{mk\pm L} - P_{mk\pm R},~~
P_{mk\pm}^{(\nu)} \equiv P_{mk\pm L} + P_{mk\pm R},
\label{PMk}
\end{eqnarray}
respectively.
By the insertion of $(-1)^{l_1+l_2}$,
we obtain $\sharp ({}_{3}C_{l_1}, {}_{2}C_{l_2})_{L(R)}$
for $l_1 + l_2=$ even integer
and $-\sharp ({}_{3}C_{l_1}, {}_{2}C_{l_2})_{L(R)}$
for $l_1 + l_2=$ odd integer.
Although the above formulae (\ref{nd-ZM}) -- (\ref{nq-ZM})
are derived with no consideration for the Wilson line phases,
they still hold for the case with
non-vanishing Wilson line phases
relating to extra gauge symmetries, thanks to a hidden 
quantum-mechanical supersymmetry~\cite{GK&M2}.

We explain how the $Z_m$ elements $\mathcal{P}^{(k)}_{a \pm}$
of multiplets in $\left({}_{3}C_{l_1}, {}_{2}C_{l_2}, \cdots, {}_{p_n}C_{l_n}\right)$
decomposed from $\Psi_{\pm}$ in $[N, k] (= {}_{N}C_{k})$
are determined by the intrinsic $Z_m$ elements $\eta^{(k)}_{a \pm}$
and the representation matrices $U_a$ 
for the fundamental representation $\bm{N}=[N,1]$.
Here, $\Psi_{\pm}$ are six-dimensional Weyl fermions in $[N, k]$,
and those boundary conditions are specified by 
representation matrices $T_{\Psi\pm}[U_a, \eta^{(k)}_{a \pm}]$,
which are factorized into
$T_{\Psi\pm}[U_a, \eta^{(k)}_{a \pm}] 
= \eta^{(k)}_{a \pm} \tilde{T}_{\Psi\pm}[U_a]$,
using overall factors $\eta^{(k)}_{a \pm}$ intrinsic to fields
and ${}_{N}C_{k} \times {}_{N}C_{k}$ matrices $\tilde{T}_{\Psi\pm}[U_a]$.
Because $\mathcal{P}^{(k)}_{a \pm}$ are obtained
as eigenvalues of $T_{\Psi\pm}[U_a, \eta^{(k)}_{a \pm}]$,
we need how $T_{\Psi\pm}[U_a, \eta^{(k)}_{a \pm}]$
act multiplets 
in $\left({}_{3}C_{l_1}, {}_{2}C_{l_2}, \cdots, {}_{p_n}C_{l_n}\right)$.
The components of $\Psi_{\pm}$ are written in the form
of the antisymmetrization of $k$-ple product of 
$\bm{N}$ such as
$\displaystyle{[N, k] = ({\bm{N} \times \dots 
\times \bm{N}})_{\mbox{\tiny{A}}}}$
where a tiny subscript A means the antisymmetrization,
and the operation of $T_{\Psi\pm}[U_a, \eta^{(k)}_{a \pm}]$ on $[N, k]$
is given by $\displaystyle{\eta^{(k)}_{a\pm} ((U_a {\bm{N}}) \times 
\dots \times (U_a {\bm{N}}))_{\mbox{\tiny{A}}}}$.
We consider a simple example of a $Z_2$ element with 
$U_0 = {\mbox{diag}}([+1]_{p_1},[-1]_{p_2})$
where $[\pm 1]_{p_i}$ represents $\pm 1$ for all $p_i$ elements.
Then the $[N, k]$ of $SU(N)$ is decomposed 
into a sum of multiplets of $SU(p_1) \times SU(p_2)$ as
$\displaystyle{[N, k] = \sum_{l_1 =0}^{k}  
\left({}_{p_1}C_{l_1}, {}_{p_2}C_{l_2}\right)}$
where $N=p_1+p_2$ and $k=l_1 + l_2$.
From the observation that $\left({}_{p_1}C_{l_1}, {}_{p_2}C_{l_2}\right)$
is multiplied by $+1$ $l_1$ times
and multiplied by $-1$ $l_2$ times
through the operation of $T_{\Psi\pm}[U_0, \eta^{(k)}_{0 \pm}]$ on $[N, k]$,
we see the $Z_2$ element of $\left({}_{p_1}C_{l_1}, {}_{p_2}C_{l_2}\right)$ as
$\mathcal{P}^{(k)}_{0 \pm}
= \eta^{(k)}_{0 \pm} (+1)^{l_1} (-1)^{l_2}
= (-1)^{l_1-k} \eta^{(k)}_{0 \pm}$
where we use $k=l_1 + l_2$ and $(-1)^n = (-1)^{-n}$ ($n$ is an integer).
In this way, if $\eta^{(k)}_{a \pm}$ and $U_a$ are given,
$\mathcal{P}^{(k)}_{a \pm}$ are determined for each multiplet,
as will be done below.

We take the representation matrices for $T^2/Z_2$,
\begin{eqnarray}
&~& U_0 
= {\mbox{diag}}([+1]_{p_1},[+1]_{p_2},[+1]_{p_3},[+1]_{p_4},[-1]_{p_5},[-1]_{p_6},
[-1]_{p_7},[-1]_{p_8}),
\nonumber \\
&~& U_1 = {\mbox{diag}}([+1]_{p_1},[+1]_{p_2},[-1]_{p_3},[-1]_{p_4},
[+1]_{p_5},[+1]_{p_6},[-1]_{p_7},[-1]_{p_8}),
\nonumber \\
&~& U_2 = {\mbox{diag}}([+1]_{p_1},[-1]_{p_2},[+1]_{p_3},[-1]_{p_4},
[+1]_{p_5},[-1]_{p_6},[+1]_{p_7},[-1]_{p_8}),
\label{Z2-U}
\end{eqnarray}
where $[\pm 1]_{p_i}$ represents $\pm 1$ for all $p_i$ elements.
Then, the $Z_2$ elements $\mathcal{P}^{(k)}_{a \pm}$ of 
$\left({}_{3}C_{l_1}, {}_{2}C_{l_2}, \cdots, {}_{p_n}C_{l_n}\right)$
are determined as
\begin{eqnarray}
\mathcal{P}^{(k)}_{0 \pm}
= (-1)^{l_1+l_2+l_3+l_4 -k}\eta^{(k)}_{0 \pm},~~ 
\mathcal{P}^{(k)}_{1 \pm} 
= (-1)^{l_1+l_2+l_5+l_6-k}\eta^{(k)}_{1 \pm},~~
\mathcal{P}^{(k)}_{2 \pm}  
= (-1)^{l_1+l_3+l_5+l_7-k}\eta^{(k)}_{2 \pm}.
\label{Z2-ele}
\end{eqnarray}
In the same way, with the representation matrices for $T^2/Z_3$,
\begin{eqnarray}
&~& U_0 = {\mbox{diag}}([1]_{p_1},[1]_{p_2},[1]_{p_3},
[\omega]_{p_4},[\omega]_{p_5},[\omega]_{p_6},
[\overline{\omega}]_{p_7},[\overline{\omega}]_{p_8},
[\overline{\omega}]_{p_9}),
\nonumber \\
&~& U_1 = {\mbox{diag}}([1]_{p_1},[\omega]_{p_2},
[\overline{\omega}]_{p_3},
[1]_{p_4},[\omega]_{p_5},[\overline{\omega}]_{p_6},
[1]_{p_7},[{\omega}]_{p_8},[\overline{\omega}]_{p_9}),
\label{Z3-U}
\end{eqnarray}
we obtain relations:
\begin{eqnarray}
\mathcal{P}^{(k)}_{0 \pm}
= \omega^{l_1+l_2+l_3+2(l_4 + l_5 + l_6) -k}\eta^{(k)}_{0 \pm},~~ 
\mathcal{P}^{(k)}_{1 \pm} 
= \omega^{l_1+l_4+l_7+2(l_2 + l_5 + l_8) -k}\eta^{(k)}_{1 \pm}.
\label{Z3-ele}
\end{eqnarray}
With the representation matrices for $T^2/Z_4$,
\begin{eqnarray}
&~& U_0 = {\mbox{diag}}([+1]_{p_1},[+1]_{p_2},[+i]_{p_3},[+i]_{p_4},
[-1]_{p_5},[-1]_{p_6},[-i]_{p_7},[-i]_{p_8}),
\nonumber \\
&~& U_1 = {\mbox{diag}}([+1]_{p_1},[-1]_{p_2},[-i]_{p_3},[+i]_{p_4},
[-1]_{p_5},[+1]_{p_6},[+i]_{p_7},[-i]_{p_8}),
\label{Z4-U}
\end{eqnarray}
we obtain relations:
\begin{eqnarray}
\mathcal{P}^{(k)}_{0 \pm}
= i^{l_1+l_2+2(l_3+l_4) + 3(l_5 + l_6) -k}\eta^{(k)}_{0 \pm},~~ 
\mathcal{P}^{(k)}_{1 \pm} 
= i^{l_1+l_6+2(l_4+l_7) + 3(l_2 + l_5) -k}\eta^{(k)}_{1 \pm}.
\label{Z4-ele}
\end{eqnarray}
With the representation matrix for $T^2/Z_6$,
\begin{eqnarray}
&~& U_0  
= {\mbox{diag}}([1]_{p_1},[\varphi]_{p_2},
[\varphi^2]_{p_3},[\varphi^3]_{p_4},[\varphi^4]_{p_5},
[\varphi^5]_{p_{6}}),
\label{Z6-U}
\end{eqnarray}
we obtain relations:
\begin{eqnarray}
\mathcal{P}^{(k)}_{0 \pm}
= \varphi^{l_1+2 l_2+3 l_3+ 4 l_4 + 5 l_5 -k}\eta^{(k)}_{0 \pm}.
\label{Z6-ele}
\end{eqnarray}
The subscripts $L$ and $R$ 
on the intrinsic $Z_m$ elements are omitted
in (\ref{Z2-ele}),  (\ref{Z3-ele}), (\ref{Z4-ele}) and (\ref{Z6-ele}).
When we use ones with $L$ or $R$,
$\eta^{(k)}_{a \pm R}$ are determined from $\eta^{(k)}_{a \pm L}$ as
\begin{eqnarray}
\eta^{(k)}_{a + R} = {\rho} \eta^{(k)}_{a + L},~~
\eta^{(k)}_{a - R} = \overline{\rho} \eta^{(k)}_{a - L},
\label{etaR}
\end{eqnarray}
as seen from (\ref{rhopsi}).
Intrinsic $Z_m$ elements satisfy the consistency conditions
such as (\ref{Z2ele-Rel}),  (\ref{Z3ele-Rel})
and the corresponding ones for $T^2/Z_4$ and $T^2/Z_6$.
Hence the product of $\eta^{(k)}_{0 \pm}$ and $\eta^{(k)}_{1 \pm}$
should be $1$ or $-1$ for $T^2/Z_4$.

In the appendix C, we give formulae for flavor numbers
from a fermion in $\overline{[N, k]}(=[N, N-k])$ 
and study the relationship
between flavor numbers
from a fermion in $\overline{[N, k]}$ 
and those from a fermion in $[N, k]$
from the viewpoint of charge conjugation.

\section{Orbifold family unification 
using vectorlike representation}

Now, we study whether or not
three families of the SM matter multiplets
are derived from a massless six-dimensional Dirac fermion
(or a pair of six-dimensional Weyl fermions)  
in a vectorlike representation $[N, k] + [N, N-k]$ of $SU(N)$
($N = 8, 9$), through the orbifold breaking mechanism.

First, we explain that complete three SM families
cannot be derived from a Dirac fermion
in $[N, 1] + [N, N-1]$ or $[N, 2] + [N, N-2]$ of $SU(N)$
in our setup given in the previous section.
After the breakdown of $SU(N)$,
$d_R$ and $(l_L)^c$ can appear from
a Dirac fermion in $[N, 1]$ and 
$(d_R)^c$ and $l_L$ can appear from a Dirac fermion in 
$[N, N-1]$, but $q_L$, $(u_R)^c$ and $(e_R)^c$
cannot come from them.
In the same way, after the breakdown of $SU(N)$,
a Dirac fermion in $[N, 2]$ only generates 
one $q_L$, one $(u_R)^c$ and/or one $(e_R)^c$ at most,
and that in $[N, N-2]$ only generates 
one $(q_L)^c$, one $u_R$ and/or one $e_R$ at most.
Hence, a Dirac fermion in $[N, 3] + [N, N-3]$
has smallest components
among a possible candidate that produces 
complete three SM families.

Second, we present total numbers of models 
with the three SM families,
which originate from a Dirac fermion 
in $[N, 3] + [N, N-3]$ of $SU(8)$
and $SU(9)$.
They are summarized in Table \ref{Table:Total}.
\begin{table}[htbp]
\caption{Total numbers of models with the three families of 
the SM multiplets.}
\label{Table:Total}
\begin{center}
\begin{tabular}{c|c|c|c|c|c}
\hline                              
$SU(N)$ & Representations  
&$T^2/Z_2$&$T^2/Z_3$&$T^2/Z_4$&$T^2/Z_6$\\
\hline 
$SU(8)$ & $[8, 3]+[8, 5]$ 
& 0 (0) & 336 (4) & 56 (0) & 0(0)\\
$SU(9)$ & $[9, 3]+[9, 6]$ 
& 1152 (768) & 1188 (600) & 512 (416) & 0(0)\\
\hline
\end{tabular}
\end{center}
\end{table}
In Table \ref{Table:Total}, 
the figures in parentheses represent
numbers of models with three or more than three neutrino singlets.
We list numbers $p_i$ ($i=1, \cdots, 9$)
specifying representation matrices $U_a$
and the intrinsic $Z_3$ elements,
to derive both the three families of 
the SM multiplets and three neutrino singlets
from a fermion in $[8, 3]+[8, 5]$ 
of $SU(8)$ on $M^4\times T^2/Z_3$, 
in Table \ref{Table:threeSM+nu}.
\begin{table}[htbp]
\caption{Models with the three families of 
the SM multiplets and three neutrino singlets
from a fermion in $[8, 3]+[8, 5]$ on $M^4\times T^2/Z_3$.}
\label{Table:threeSM+nu}
\begin{center}
\begin{tabular}{c|c|c|c|c}
\hline                              
$(p_1, p_2, p_3, p_4, p_5, p_6, p_7, p_8, p_9)$ 
& $(\eta_{0+L}^{(3)}, \eta_{1+L}^{(3)})$ 
& $(\eta_{0-L}^{(3)}, \eta_{1-L}^{(3)})$
& $(\eta_{0+L}^{(5)}, \eta_{1+L}^{(5)})$ 
& $(\eta_{0-L}^{(5)}, \eta_{1-L}^{(5)})$\\
\hline
$(3, 2, 0, 0, 1, 1, 1, 0, 0)$
& $(\omega^2, 1)$ & $(\omega, \omega^2)$ 
& $(\omega^2, 1)$ & $(\omega, \omega^2)$ \\
$(3, 2, 0, 0, 1, 1, 1, 0, 0)$
& $(\omega^2, 1)$ & $(\omega^2, 1)$ 
& $(\omega^2, 1)$ & $(\omega^2, 1)$ \\
$(3, 2, 0, 0, 0, 1, 1, 0, 1)$
& $(\omega^2, 1)$ & $(\omega, \omega^2)$ 
& $(\omega^2, 1)$ & $(\omega, \omega^2)$ \\
$(3, 2, 0, 0, 0, 1, 1, 0, 1)$
& $(\omega, \omega^2)$ & $(\omega, \omega^2)$ 
& $(\omega, \omega^2)$ & $(\omega, \omega^2)$ \\ \hline
\end{tabular}
\end{center}
\end{table}
In Table \ref{Table:threeSM+nu},
only the intrinsic $Z_3$ elements for the $\psi_{\pm L}$
are written, and those for the $\psi_{\pm R}$
can be seen from (\ref{etaR}).

Third, we give examples concerning 
the appearance of three SM families,
using the first and second models in Table \ref{Table:threeSM+nu}.
By taking $(p_1, p_2, p_3, p_4, p_5, p_6, p_7, p_8, p_9)
=(3, 2, 0, 0, 1, 1, 1, 0, 0)$,
the $SU(8)$ gauge symmetry is broken down as 
\begin{eqnarray}
SU(8) \rightarrow SU(3)_C \times SU(2)_L \times U(1)^4.
\label{4.SU8}
\end{eqnarray}
Note that the residual gauge symmetry
does not contain any non-Abelian continuous flavor symmetry.
Then, {\bf 56}($=[8, 3]$) and $\overline{{\bf 56}}$($=[8, 5]$)
are decomposed into
particles with the SM gauge quantum numbers
and its opposite ones, as shown in Table \ref{Table:[8,3]}
and Table \ref{Table:[8,5]}, respectively.
\begingroup
\renewcommand{\arraystretch}{1.2}
\begin{table}[htbp]
\begin{center}
\caption{Decomposition of {\bf 56}
for $(p_1, p_2, p_3, p_4, p_5, p_6, p_7, p_8, p_9)
=(3, 2, 0, 0, 1, 1, 1, 0, 0)$.}
\label{Table:[8,3]}
~~\\
\begin{tabular}{c|c|c|c}
\hline
$\psi_{\pm L}^{[8,3]}$&$\psi_{\pm R}^{[8,3]}$
&$(l_1, l_2, l_5, l_6, l_7)$ & 
$(\mathcal{P}_{0\pm}^{(3)}, \mathcal{P}_{1\pm}^{(3)})$ \\
\hline \hline
$(e_R^{\prime})^c$ & $e_R$ 
& $(3, 0, 0, 0, 0)$ &
$(\eta_{0\pm}^{(3)}, \eta_{1\pm}^{(3)})$ \\
\hline
$q_L^{\prime}$ & $(q_L)^c$
&$(2, 1, 0, 0, 0)$ &
$(\eta_{0\pm}^{(3)}, \omega \eta_{1\pm}^{(3)})$ \\
\hline
$(u_R^{\prime})^c$ & $u_R$
&$(1, 2, 0, 0, 0)$ &
$(\eta_{0\pm}^{(3)}, \omega^2 \eta_{1\pm}^{(3)})$ \\
\hline
 & &$(2, 0, 1, 0, 0)$ &
$(\omega \eta_{0\pm}^{(3)}, \omega \eta_{1\pm}^{(3)})$ \\
$(u_R)^c$ & $u_R^{\prime}$
&$(2, 0, 0, 1, 0)$ &
$(\omega \eta_{0\pm}^{(3)}, \omega^2 \eta_{1\pm}^{(3)})$ \\
& &$(2, 0, 0, 0, 1)$ &
$(\omega^2 \eta_{0\pm}^{(3)}, \eta_{1\pm}^{(3)})$\\
\hline
 & &$(1, 1, 1, 0, 0)$ &
$(\omega \eta_{0\pm}^{(3)}, \omega^2 \eta_{1\pm}^{(3)})$ \\
$q_L$ & $(q_L^{\prime})^c$
&$(1, 1, 0, 1, 0)$ &
$(\omega \eta_{0\pm}^{(3)}, \eta_{1\pm}^{(3)})$ \\
 & &$(1, 1, 0, 0, 1)$ &
$(\omega^2 \eta_{0\pm}^{(3)}, \omega \eta_{1\pm}^{(3)})$ \\
\hline
 & &$(0, 2, 1, 0, 0)$ &
$(\omega \eta_{0\pm}^{(3)}, \eta_{1\pm}^{(3)})$ \\
$(e_R)^c$ & $e_R^{\prime}$
&$(0, 2, 0, 1, 0)$ &
$(\omega \eta_{0\pm}^{(3)}, \omega \eta_{1\pm}^{(3)})$ \\
& &$(0, 2, 0, 0, 1)$ &
$(\omega^2 \eta_{0\pm}^{(3)}, \omega^2 \eta_{1\pm}^{(3)})$\\
\hline
 & &$(1, 0, 1, 1, 0)$ &
$(\omega^2 \eta_{0\pm}^{(3)}, \eta_{1\pm}^{(3)})$ \\
$(d_R^{\prime})^c$&$d_R$
&$(1, 0, 1, 0, 1)$ &
$(\eta_{0\pm}^{(3)}, \omega \eta_{1\pm}^{(3)})$ \\
 & &$(1, 0, 0, 1, 1)$ &
$(\eta_{0\pm}^{(3)}, \omega^2 \eta_{1\pm}^{(3)})$ \\
\hline
 & &$(0, 1, 1, 1, 0)$ &
$(\omega^2 \eta_{0\pm}^{(3)}, \omega \eta_{1\pm}^{(3)})$ \\
$l_L^{\prime}$&$(l_L)^c$
&$(0, 1, 1, 0, 1)$ &
$(\eta_{0\pm}^{(3)}, \omega^2 \eta_{1\pm}^{(3)})$ \\
 & &$(0, 1, 0, 1, 1)$ &
$(\eta_{0\pm}^{(3)}, \eta_{1\pm}^{(3)})$ \\
\hline
$(\nu_R)^c$&${\nu}_R$
&$(0, 0, 1, 1, 1)$ &
$(\omega \eta_{0\pm}^{(3)}, \eta_{1\pm}^{(3)})$ \\
\hline
\end{tabular}
\end{center}
\end{table}
\endgroup
In the first and second columns, 
particles are denoted by using the symbols in the SM,
and those with primes are regarded as mirror particles,
which are particles with opposite quantum numbers under 
the SM gauge group.
In the third column, $l_i$ not on the list are zero.
In the fourth column, the subscripts $L$ and $R$
are omitted on the intrinsic $Z_3$ elements.
\begingroup
\renewcommand{\arraystretch}{1.2}
\begin{table}[htbp]
\begin{center}
\caption{Decomposition of $\overline{\bf 56}$
for $(p_1, p_2, p_3, p_4, p_5, p_6, p_7, p_8, p_9)
=(3, 2, 0, 0, 1, 1, 1, 0, 0)$.
}
\label{Table:[8,5]}
~~\\
\begin{tabular}{c|c|c|c}
\hline
$\psi_{\pm L}^{[8,5]}$&$\psi_{\pm R}^{[8,5]}$
&$(l_1, l_2, l_5, l_6, l_7)$ & 
$(\mathcal{P}_{0\pm}^{(5)}, \mathcal{P}_{1\pm}^{(5)})$ \\
\hline \hline
& & $(3, 0, 1, 1, 0)$ &
$(\omega^2\eta_{0\pm}^{(5)}, \eta_{1\pm}^{(5)})$ \\
$(e_R^{\prime})^c$ & $e_R$ 
& $(3, 0, 1, 0, 1)$ &
$(\eta_{0\pm}^{(5)}, \omega\eta_{1\pm}^{(5)})$ \\
& & $(3, 0, 0, 1, 1)$ &
$(\eta_{0\pm}^{(5)}, \omega^2 \eta_{1\pm}^{(5)})$ \\
\hline
& & $(2, 1, 1, 1, 0)$ &
$(\omega^2 \eta_{0\pm}^{(5)}, \omega \eta_{1\pm}^{(5)})$ \\
$q_L^{\prime}$ & $(q_L)^c$
&$(2, 1, 1, 0, 1)$ &
$(\eta_{0\pm}^{(5)}, \omega^2 \eta_{1\pm}^{(5)})$ \\
& & $(2, 1, 0, 1, 1)$ &
$(\eta_{0\pm}^{(5)}, \eta_{1\pm}^{(5)})$ \\
\hline
& & $(1, 2, 1, 1, 0)$ &
$(\omega^2 \eta_{0\pm}^{(5)}, \omega^2 \eta_{1\pm}^{(5)})$ \\
$(u_R^{\prime})^c$ & $u_R$
&$(1, 2, 1, 0, 1)$ &
$(\eta_{0\pm}^{(5)}, \eta_{1\pm}^{(5)})$ \\
& &$(1, 2, 0, 1, 1)$ &
$(\eta_{0\pm}^{(5)}, \omega \eta_{1\pm}^{(5)})$ \\
\hline
$(u_R)^c$ & $u_R^{\prime}$
&$(2, 0, 1, 1, 1)$ &
$(\omega \eta_{0\pm}^{(5)}, \eta_{1\pm}^{(5)})$ \\
\hline
$q_L$ & $(q_L^{\prime})^c$
&$(1, 1, 1, 1, 1)$ &
$(\omega \eta_{0\pm}^{(5)}, \omega \eta_{1\pm}^{(5)})$ \\
\hline
$(e_R)^c$ & $e_R^{\prime}$
&$(0, 2, 1, 1, 1)$ &
$(\omega \eta_{0\pm}^{(5)}, \omega^2 \eta_{1\pm}^{(5)})$ \\
\hline
 & &$(2, 2, 1, 0, 0)$ &
$(\omega \eta_{0\pm}^{(5)}, \eta_{1\pm}^{(5)})$ \\
$(d_R)^c$&$d_R^{\prime}$
&$(2, 2, 0, 1, 0)$ &
$(\omega \eta_{0\pm}^{(5)}, \omega \eta_{1\pm}^{(5)})$ \\
 & &$(2, 2, 0, 0, 1)$ &
$(\omega^2 \eta_{0\pm}^{(5)}, \omega^2 \eta_{1\pm}^{(5)})$ \\
\hline
 & &$(3, 1, 1, 0, 0)$ &
$(\omega \eta_{0\pm}^{(5)}, \omega^2 \eta_{1\pm}^{(5)})$ \\
$l_L$&$(l_L^{\prime})^c$
&$(3, 1, 0, 1, 0)$ &
$(\omega \eta_{0\pm}^{(5)}, \eta_{1\pm}^{(5)})$ \\
 & &$(3, 1, 0, 0, 1)$ &
$(\omega^2 \eta_{0\pm}^{(5)}, \omega \eta_{1\pm}^{(5)})$ \\
\hline
$(\nu_R)^c$&${\nu}_R$
&$(3, 2, 0, 0, 0)$ &
$(\eta_{0\pm}^{(5)}, \omega^2 \eta_{1\pm}^{(5)})$ \\
\hline
\end{tabular}
\end{center}
\end{table}
\endgroup

We give an assignment of intrinsic $Z_3$ elements 
and particle contents 
to derive three SM families
and three neutrino singlets as zero modes 
in Table \ref{Table:Z3-1}.
\begingroup
\renewcommand{\arraystretch}{1.2}
\begin{table}[htbp]
\begin{center}
\caption{The particle contents as zero modes
obtained from ${\bf 56}$ and $\overline{\bf 56}$.}
\label{Table:Z3-1}
~~\\
\begin{tabular}{c|c|c c c c c c}
\hline
multiplets & $(\eta_{0\pm}^{(k)}, \eta_{1\pm}^{(k)})$
& $(d_R)^c$ & $l_L$ & $(u_R)^c$ & $(e_R)^c$ & $q_L$ & $(\nu_R)^c$\\
\hline \hline
$\psi_{+ L}^{[8,3]}$& $(\omega^2, 1)$
& & & & $(e_R)^c$ & $q_L$ & $(\nu_R)^c$\\
\hline
$\psi_{+ R}^{[8,3]}$& $(1, \omega)$
& $d_R$ & $(l_L)^c$ & $u_R$ & & & \\
\hline
$\psi_{- L}^{[8,3]}$& $(\omega, \omega^2)$
& & $l_L^{\prime}$ & & & $q_L$ & \\
\hline
$\psi_{- R}^{[8,3]}$& $(1, \omega)$
& $d_R$ & $(l_L)^c$ & $u_R$ & & & \\
\hline
$\psi_{+ L}^{[8,5]}$& $(\omega^2, 1)$
& $(d_R)^c$ & $l_L$ & $(u_R)^c$ & & & \\
\hline
$\psi_{+ R}^{[8,5]}$& $(1, \omega)$
 & & & & $e_R$ & $(q_L)^c$ & $\nu_R$\\
\hline
$\psi_{- L}^{[8,5]}$& $(\omega, \omega^2)$
& & $l_L$ & & & $q_L^{\prime}$ & \\
\hline
$\psi_{- R}^{[8,5]}$& $(1, \omega)$
 & & & & $e_R$ & $(q_L)^c$ & $\nu_R$\\
\hline
\end{tabular}
\end{center}
\end{table}
\endgroup
As seen from Table \ref{Table:Z3-1},
just three sets of SM fermions 
$(q_L^i, (u_R^i)^c, (d_R^i)^c, l_L^i, (e_R^i)^c)$
and three kinds of neutrino singlets 
$((\nu_{R})^c$ and $\nu^R)$ 
are originated as zero modes
from $\psi_{\pm L}^{[8,3]} + \psi_{\pm R}^{[8,3]}
+ \psi_{\pm L}^{[8,5]} + \psi_{\pm R}^{[8,5]}$
with suitable intrinsic $Z_3$ elements,
after the survival hypothesis works.
Mirror particles can disappear by acquiring heavy masses,
that is, the $l_L^{\prime}$ in $\psi_{- L}^{[8,3]}$
can be massive with one of $l_L$, $(l_L)^c$ or a mixture of them
and the $q_L^{\prime}$ in $\psi_{- L}^{[8,5]}$
can be massive with one of $q_L$, $(q_L)^c$ or a mixture of them.

In the same way,
we can obtain particle contents with just three SM families
and three neutrino singlets as zero modes 
from $\psi_{\pm L}^{[8,3]} + \psi_{\pm R}^{[8,3]}
+ \psi_{\pm L}^{[8,5]} + \psi_{\pm R}^{[8,5]}$
with intrinsic $Z_3$ elements
assigned in Table \ref{Table:Z3-2},
after the survival hypothesis works.
\begingroup
\renewcommand{\arraystretch}{1.2}
\begin{table}[htbp]
\begin{center}
\caption{Another assignment of intrinsic $Z_3$ elements
and the particle contents as zero modes
obtained from ${\bf 56}$ and $\overline{\bf 56}$.}
\label{Table:Z3-2}
~~\\
\begin{tabular}{c|c|c c c c c c}
\hline
multiplets & $(\eta_{0\pm}^{(k)}, \eta_{1\pm}^{(k)})$
& $(d_R)^c$ & $l_L$ & $(u_R)^c$ & $(e_R)^c$ & $q_L$ & $(\nu_R)^c$\\
\hline \hline
$\psi_{+ L}^{[8,3]}$& $(\omega^2, 1)$
& & & & $(e_R)^c$ & $q_L$ & $(\nu_R)^c$\\
\hline
$\psi_{+ R}^{[8,3]}$& $(1, \omega)$
& $d_R$ & $(l_L)^c$ & $u_R$ & & & \\
\hline
$\psi_{- L}^{[8,3]}$& $(\omega^2, 1)$
& & & & $(e_R)^c$ & $q_L$ & $(\nu_R)^c$\\
\hline
$\psi_{- R}^{[8,3]}$& $(\omega, \omega^2)$
& & $(l_L)^c$ & & & $(q_L^{\prime})^c$ & \\
\hline
$\psi_{+ L}^{[8,5]}$& $(\omega^2, 1)$
& $(d_R)^c$ & $l_L$ & $(u_R)^c$ & & & \\
\hline
$\psi_{+ R}^{[8,5]}$& $(1, \omega)$
 & & & & $e_R$ & $(q_L)^c$ & $\nu_R$\\
\hline
$\psi_{- L}^{[8,5]}$& $(\omega^2, 1)$
& $(d_R)^c$ & $l_L$ & $(u_R)^c$ & & & \\
\hline
$\psi_{- R}^{[8,5]}$& $(\omega, \omega^2)$
& & $(l_L^{\prime})^c$ & & & $(q_L)^c$ & \\
\hline
\end{tabular}
\end{center}
\end{table}
\endgroup

Finally, we point out that the classification of
our models has not yet been completed in our setup.
Concretely, we consider the breaking pattern (\ref{GSB})
with the identification of $SU(p_1) = SU(3)_C$ and $SU(p_2) = SU(2)_L$,
and take the diagonal representation matrices (\ref{Z2-U}),
(\ref{Z3-U}), (\ref{Z4-U}) and (\ref{Z6-U}).
Based on the representation matrices given above, 
there is a choice to take $p_i = 3$ and $p_j = 2$
with $(i, j) \ne (1, 2)$ as $SU(3)_C \times SU(2)_L$.
Or provided that $p_1 = 3$ and $p_2 = 2$,
we can choose different diagonal representation matrices,
that are obtained by the exchange of components 
in the above ones.
Same results are obtained from most of them, but
there are independent choices 
to generate models different from those mentioned in this section.
Complete analysis and classification will be reported,
including results from a fermion in $[N, k] + [N, N-k]$ ($k \ge 4$),
in a forthcoming paper~\cite{G&K2}.

\section{Conclusions}

We have studied the possibility of family unification 
on the basis of $SU(N)$ gauge theory 
on the six-dimensional space-time
$M^4\times T^2/Z_m$ ($m=2, 3, 4, 6$).
We have obtained enormous numbers of models 
with three families of the SM matter multiplets
are derived from a massless six-dimensional Dirac fermion 
in a vectorlike representation $[N, 3] + [N, N-3]$ of $SU(N)$
($N = 8, 9$), through the orbifold breaking mechanism,
and found
models with three or more than three neutrino singlets
and without any non-Abelian continuous flavor gauge symmetries.
We have shown a feature
that each flavor number from a fermion in $[N, k]$ 
with intrinsic $Z_m$ elements $\eta^{(k)}_{a \pm}$
is equal to that from a fermion in $\overline{[N, k]}(=[N,N-k])$ 
with appropriate $\eta^{(N-k)}_{a \pm}$,
because there is a one-to-one 
correspondence between
zero modes from a Weyl fermion in
$[N,k]$ with $\eta^{(k)}_{a \pm}$
and those from a Weyl fermion in $[N,N-k]$ with
appropriate $\eta^{(N-k)}_{a \pm}$,
using the equivalence under the charge conjugation.

Now, we have several problems as a future work.

It is meaningful to study phenomenological implications
relating to the breakdown of extra $U(1)$ gauge symmetries,
$D$-term contributions to scalar (squark, slepton and Higgs)
masses and
the generation of realistic fermion masses
and family mixing,
based on $SU(8)$ models illustrated in Sect. 4.
The $SU(8)$ models are attractive,
because there is no non-Abelian continuous gauge group,
and extra $U(1)$ gauge bosons can be massive
by the vacuum expectation values of the SM singlets scalar fields.
Moreover, superpartners of neutrino singlets can be 
candidates of such scalar fields.
In SUSY models, there appear $D$-term contributions 
to scalar masses 
after the breakdown of extra gauge symmetries,
if soft SUSY breaking terms have a non-universal structure,
and its contributions lift 
the mass degeneracy~\cite{D,H&K,KM&Y1,KM&Y2,K&T}.
Under assumptions that SUSY is broken down by
the dynamics on a brane and non-universal soft SUSY breaking
terms are induced,
the $D$-term contributions have been studied 
in the framework of $SU(N)$ orbifold GUTs~\cite{K&K1,K&K2,KK&M},
and they can become useful probes to specify
a realistic model in GUTs.
Then we need to reconsider the anomaly cancellations 
on a construction of SUSY models,
because various fermions exist there.
Fermion mass hierarchy and family mixing can occur 
through the Froggatt-Nielsen mechanism~\cite{FN}
on the breakdown of extra $U(1)$ gauge symmetries 
and/or the suppression of brane-localized Yukawa coupling constants 
among brane weak Higgs doublets and bulk fermions 
with the volume suppression factor~\cite{Y}.

It would be interesting to reconstruct our models 
in the framework of $E_8$ gauge theory or superstring theory.
Various 4-dimensional string models including three families 
have been constructed from several methods,
see e.g.~\cite{I&U} and references therein 
for useful articles.\footnote{
See also Ref.~\cite{M&Y,M} and references therein for recent works.
}
It has been pointed out that $SO(1,D-1)$ space-time symmetry 
can lead to family structure~\cite{B&N1,B&N2},
and hence it would offer a hint to explore 
the family structure in our models. 

Furthermore, it would be intriguing to study 
cosmological implications of the class of models presented in this paper, 
see e.g.~\cite{Khlopov:1999rs} and references therein 
for useful articles toward this direction.

\section*{Acknowledgements}
This work was supported in part by scientific grants 
from Iwanami Fu-Jukai and the MEXT-Supported Program 
for the Strategic Research Foundation at Private Universities 
“Topological Science” under Grant No.~S1511006 (Y.~G.)
and from the Ministry of Education, Culture,
Sports, Science and Technology under Grant No.~17K05413 (Y.~K.).

\appendix

\section{$T^2/Z_m$ orbifold}
\label{app-A}

\subsection{$T^2/Z_2$}
\label{app-A1}

\begin{figure}[ht!]
\caption{Orbifold $T^2/Z_2$}
\label{F1}
\begin{center}
\unitlength 0.1in
\begin{picture}( 15.0000, 15.7000)(  3.6000,-17.5000)
%
\special{pn 13}%
\special{pa 620 1616}%
\special{pa 1790 1616}%
\special{fp}%
\special{sh 1}%
\special{pa 1790 1616}%
\special{pa 1724 1596}%
\special{pa 1738 1616}%
\special{pa 1724 1636}%
\special{pa 1790 1616}%
\special{fp}%
%
\special{pn 13}%
\special{pa 600 1620}%
\special{pa 614 450}%
\special{fp}%
\special{sh 1}%
\special{pa 614 450}%
\special{pa 592 516}%
\special{pa 612 504}%
\special{pa 632 518}%
\special{pa 614 450}%
\special{fp}%
%
\special{pn 8}%
\special{pa 600 450}%
\special{pa 1790 450}%
\special{dt 0.045}%
%
\special{pn 8}%
\special{pa 1790 450}%
\special{pa 1790 1620}%
\special{dt 0.045}%
%
\special{pn 20}%
\special{sh 1}%
\special{ar 610 1620 10 10 0  6.28318530717959E+0000}%
\special{sh 1}%
\special{ar 600 1620 10 10 0  6.28318530717959E+0000}%
%
\special{pn 20}%
\special{sh 1}%
\special{ar 620 1010 10 10 0  6.28318530717959E+0000}%
\special{sh 1}%
\special{ar 610 1010 10 10 0  6.28318530717959E+0000}%
%
\special{pn 20}%
\special{sh 1}%
\special{ar 1190 1010 10 10 0  6.28318530717959E+0000}%
\special{sh 1}%
\special{ar 1180 1010 10 10 0  6.28318530717959E+0000}%
%
\special{pn 20}%
\special{sh 1}%
\special{ar 1200 1620 10 10 0  6.28318530717959E+0000}%
\special{sh 1}%
\special{ar 1190 1620 10 10 0  6.28318530717959E+0000}%
\put(4.4000,-18.4000){\makebox(0,0)[lb]{O}}%
\put(18.6000,-16.6000){\makebox(0,0)[lb]{$e_1$}}%
\put(5.3000,-3.5000){\makebox(0,0)[lb]{$e_2$}}%
\put(3.6000,-11.0000){\makebox(0,0)[lb]{$\frac{e_2}{2}$}}%
\put(11.3000,-19.2000){\makebox(0,0)[lb]{$\frac{e_1}{2}$}}%
\put(11.9000,-9.4000){\makebox(0,0)[lb]{$\frac{e_1+e_2}{2}$}}%
\end{picture}%
\end{center}
\end{figure}
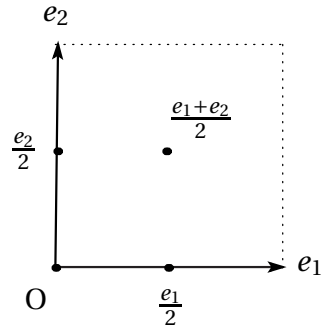

The orbifold $T^2/Z_2$ is obtained by identifying $z + e_1$, $z + e_2$
and $-z$ with $z$.
Here $e_1 = 1$ and $e_2 = i$.
The resultant space is depicted in Figure \ref{F1}.
Fix points $z_{\text{fp}}$ satisfy
$z_{\text{fp}} = -z_{\text{fp}} + a e_1 + b e_2$
where $a$ and $b$ are integers.
There are four kinds of fixed points $0$, $e_1/2$, $e_2/2$, $(e_1+e_2)/ 2$.
Around these points, we define six kinds of transformations:
\begin{eqnarray}
&~& s_0: z\rightarrow -z,~~
s_1: z\rightarrow -z+e_1, ~~
s_2: z\rightarrow -z+e_2, ~~
s_3: z\rightarrow -z+e_1+e_2, ~~
\nonumber \\
&~& t_1: z\rightarrow z+e_1,~~
t_2: z\rightarrow z+e_2
\label{Z2-transf}
\end{eqnarray}
and they satisfy the relations:
\begin{eqnarray}
&~& s_0^2=s_1^2=s_2^2=s_3^2=I,~s_1=t_1s_0,~s_2=t_2s_0,
\nonumber \\
&~& s_3=t_1t_2s_0=s_1s_0s_2=s_2s_0s_1,~t_1t_2=t_2t_1,
\label{Z2-relations}
\end{eqnarray}
where $I$ is the identity operation.

The boundary conditions of six-dimensional bulk fields are specified 
by representation matrices $(U_0, U_1, U_2, U_3, V_1, V_2)$
and intrinsic $Z_2$ elements $(\eta_0, \eta_1, \eta_2, \eta_3, \xi_1, \xi_2)$
corresponding to the above transformations.
These matrices and $Z_2$ elements satisfy the relations: 
\begin{eqnarray}
&~& U_{0}^2=U_{1}^2=U_{2}^2=U_3^2=I,~~
U_{1}=V_1U_{0},~~ U_{2}=V_2U_{0},
\nonumber \\
&~& U_3=V_1V_2U_0=U_{1}U_0U_{2}
=U_{2}U_0U_{1},~~ V_1V_2=V_2V_1,
\label{Z2-Rel}\\
&~& \eta_{0}^2=\eta_{1}^2=\eta_{2}^2=\eta_3^2=1,~~
\eta_{1}=\xi_1 \eta_{0},~~ \eta_{2}=\xi_2 \eta_{0},~~
\eta_3=\xi_1 \xi_2 \eta_0=\eta_{1} \eta_0 \eta_{2},
\label{Z2ele-Rel}
\end{eqnarray}
as the consistency conditions.
Here, we omit the subscripts specifying fields
and/or chiralities such as $\Phi$, $\pm$, $L$ and/or $R$.
Note that $\eta_{1} \eta_0 \eta_{2}=\eta_{2} \eta_0 \eta_{1}$ and $\xi_1 \xi_2=\xi_2 \xi_1$
hold automatically because intrinsic $Z_m$ elements are numbers.
From (\ref{Z2-relations}) and (\ref{Z2-Rel}),
we find that any three transformations are independent
and others are constructed as combinations of them.
We choose the transformations $s_0:z \to -z$, $s_1:z \to 1-z$
and $s_2:z \to i-z$ and the corresponding matrices 
$U_0$, $U_1$ and $U_2$.

\subsection{$T^2/Z_3$}
\label{app-A2}

\begin{figure}[ht!]
\caption{Orbifold $T^2/Z_3$}
\label{F2}
\begin{center}
\unitlength 0.1in
\begin{picture}( 20.0000, 13.2000)(  8.9000,-20.3000)
%
\special{pn 13}%
\special{pa 1650 1976}%
\special{pa 2820 1976}%
\special{fp}%
\special{sh 1}%
\special{pa 2820 1976}%
\special{pa 2754 1956}%
\special{pa 2768 1976}%
\special{pa 2754 1996}%
\special{pa 2820 1976}%
\special{fp}%
%
\special{pn 13}%
\special{pa 1630 1980}%
\special{pa 1026 978}%
\special{fp}%
\special{sh 1}%
\special{pa 1026 978}%
\special{pa 1044 1046}%
\special{pa 1054 1024}%
\special{pa 1078 1026}%
\special{pa 1026 978}%
\special{fp}%
%
\special{pn 8}%
\special{pa 1040 980}%
\special{pa 2230 980}%
\special{dt 0.045}%
%
\special{pn 8}%
\special{pa 2204 986}%
\special{pa 2820 1980}%
\special{dt 0.045}%
%
\special{pn 20}%
\special{sh 1}%
\special{ar 1640 1980 10 10 0  6.28318530717959E+0000}%
\special{sh 1}%
\special{ar 1630 1980 10 10 0  6.28318530717959E+0000}%
%
\special{pn 20}%
\special{sh 1}%
\special{ar 2180 1640 10 10 0  6.28318530717959E+0000}%
\special{sh 1}%
\special{ar 2170 1640 10 10 0  6.28318530717959E+0000}%
%
\special{pn 20}%
\special{sh 1}%
\special{ar 1630 1310 10 10 0  6.28318530717959E+0000}%
\special{sh 1}%
\special{ar 1620 1310 10 10 0  6.28318530717959E+0000}%
\put(14.7000,-22.0000){\makebox(0,0)[lb]{O}}%
\put(28.9000,-20.2000){\makebox(0,0)[lb]{$e_1$}}%
\put(8.9000,-8.8000){\makebox(0,0)[lb]{$e_2$}}%
\put(15.0000,-12.7000){\makebox(0,0)[lb]{$\frac{e_1+2e_2}{3}$}}%
\put(21.3000,-16.2000){\makebox(0,0)[lb]{$\frac{2e_1+e_2}{3}$}}%
\end{picture}%
\end{center}
\end{figure}
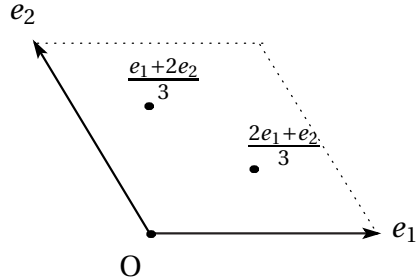

The orbifold $T^2/Z_3$ is obtained by identifying 
$z + e_1$, $z + e_2$ and $\omega z$ with $z$.
Here $e_1 =1$ and $e_2 = \omega = e^{2\pi i/3}$.
The resultant space is depicted in Figure \ref{F2}.
Fixed points satisfying 
$z_{\text{fp}} = \omega z_{\text{fp}} + a e_1 + b e_2$ ($a$, $b$:integers)
are $z=0$, $(2e_1 + e_2)/3$ and $(e_1 + 2 e_2)/3$.
Around these points, we define five kinds of transformations:
\begin{eqnarray}
&~& s_{0}: z \to \omega z,~~
s_{1}: z \to \omega z+e_{1},~~ s_{2}: z \to \omega z+e_{1}+e_{2}, 
\nonumber \\
&~& t_{1}: z \to z + e_{1}, ~~ t_{2}: z \to z + e_{2}
\label{Z3-transf}
\end{eqnarray}
and they satisfy the relations:
\begin{eqnarray}
&~& s_{0}^{3}=s_{1}^{3}=s_{2}^{3}
=s_{0}s_{1}s_{2} =s_{1}s_{2}s_{0} =s_{2}s_{0}s_{1}=I , \nonumber \\
&~& s_{1}=t_{1}s_{0},~~ s_{2}=t_{2}t_{1}s_{0}, ~~ t_{1}t_{2}=t_{2}t_{1}.
\label{Z3-relations}
\end{eqnarray}
The boundary conditions of bulk fields are specified by matrices 
$(U_0, U_1, U_2, V_1, V_2)$
and intrinsic $Z_3$ elements $(\eta_0, \eta_1, \eta_2, \xi_1, \xi_2)$
satisfying the relations: 
\begin{eqnarray}
&~& U_0^3=U_1^3=U_2^3=U_0 U_1 U_2
=U_1 U_2 U_0= U_2 U_0 U_1=I,
\nonumber \\
&~& U_1= V_1 U_0,~~
U_2=V_2 V_1 U_0,~~ V_1 V_2=V_2 V_1,
\label{Z3-Rel}\\
&~& \eta_0^3=\eta_1^3=\eta_2^3=\eta_0 \eta_1 \eta_2=1,~~
\eta_1= \xi_1 \eta_0,~~
\eta_2=\xi_2 \xi_1 \eta_0,
\label{Z3ele-Rel}
\end{eqnarray}
where we omit the subscripts specifying fields
and/or chiralities such as $\Phi$, $\pm$, $L$ and/or $R$.
Because two of these matrices are independent,
we choose representation matrices $U_0$ and $U_1$
corresponding to the transformations $s_0:z \to e^{2\pi i/3}z$
and $s_1:z \to e^{2\pi i/3}z + 1$.

\subsection{$T^2/Z_4$}
\label{app-A3}

The orbifold $T^2/Z_4$ is obtained by identifying 
$z +e_1$, $z + e_2$, $iz$ and $-z$ with $z$.
Here $e_1 = 1$ and $e_2 = i$.
The resultant space is depicted as the same figure as $T^2/Z_2$.
Fixed points are $z_{\text{fp}}=0$ 
and $(e_1+e_2)/2$ for the $Z_4$ transformation $z \to iz$
and $z_{\text{fp}} =0$, $e_1/2$, $e_2/2$ and $(e_1+e_2)/2$ 
for the $Z_2$ transformation $z \to -z$.
Around these points, we define eight kinds of transformations:
\begin{eqnarray}
&~& s_0: z\rightarrow iz, ~~
s_1: z\rightarrow iz+e_1, ~~ s_{20}: z\rightarrow -z, 
\nonumber \\
&~& s_{21}: z\rightarrow -z+e_1,~~
s_{22}: z\rightarrow -z+e_2, ~~
s_{23}: z\rightarrow -z+e_1+e_2, 
\nonumber \\
&~& t_1: z\rightarrow z+e_1, ~~
t_2: z\rightarrow z+e_2
\label{Z4-transf}
\end{eqnarray}
and they satisfy the relations:
\begin{eqnarray}
&~& s_0^4=s_1^4=s_{20}^2=s_{21}^2=s_{22}^2=s_{23}^2=I,~~
s_1=t_1s_0,~~s_{21}=t_1s_{20}, \nonumber \\
&~& s_{22}=t_2s_{20},~~s_{20}=s_0^2,~~s_{21}=s_1s_0,~~s_{22}=s_0s_1, 
\nonumber \\
&~& s_{23}=t_1t_2s_{20}=s_{21}s_{20}s_{22}=s_{22}s_{20}s_{21},~~t_1t_2=t_2t_1.
\label{Z4-relations}
\end{eqnarray}
The $Z_4$ transformations $s_0$ and $s_1$ are independent of each other
and those representation
matrices are denoted as $U_0$ and $U_1$, respectively.
Other representation matrices are determined uniquely,
if $U_0$ and $U_1$ are given.

\subsection{$T^2/Z_6$}
\label{app-A4}

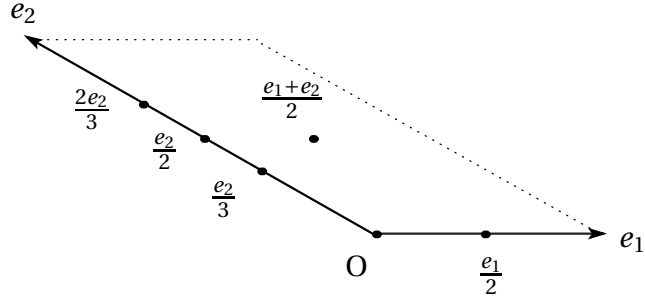
\begin{figure}[ht!]
\caption{Orbifold $T^2/Z_6$}
\label{F3}
\begin{center}
\unitlength 0.1in
\begin{picture}( 31.9000, 14.2000)(  6.8000,-25.2000)
%
\special{pn 13}%
\special{pa 2620 2376}%
\special{pa 3790 2376}%
\special{fp}%
\special{sh 1}%
\special{pa 3790 2376}%
\special{pa 3724 2356}%
\special{pa 3738 2376}%
\special{pa 3724 2396}%
\special{pa 3790 2376}%
\special{fp}%
%
\special{pn 13}%
\special{pa 2590 2380}%
\special{pa 774 1346}%
\special{fp}%
\special{sh 1}%
\special{pa 774 1346}%
\special{pa 822 1396}%
\special{pa 820 1372}%
\special{pa 842 1362}%
\special{pa 774 1346}%
\special{fp}%
%
\special{pn 8}%
\special{pa 790 1360}%
\special{pa 1980 1360}%
\special{dt 0.045}%
%
\special{pn 20}%
\special{sh 1}%
\special{ar 2610 2380 10 10 0  6.28318530717959E+0000}%
\special{sh 1}%
\special{ar 2600 2380 10 10 0  6.28318530717959E+0000}%
%
\special{pn 20}%
\special{sh 1}%
\special{ar 2280 1880 10 10 0  6.28318530717959E+0000}%
\special{sh 1}%
\special{ar 2270 1880 10 10 0  6.28318530717959E+0000}%
\put(24.4000,-26.0000){\makebox(0,0)[lb]{O}}%
\put(38.7000,-24.7000){\makebox(0,0)[lb]{$e_1$}}%
\put(6.8000,-12.7000){\makebox(0,0)[lb]{$e_2$}}%
\put(10.1000,-18.3000){\makebox(0,0)[lb]{$\frac{2e_2}{3}$}}%
\put(17.3000,-22.9000){\makebox(0,0)[lb]{$\frac{e_2}{3}$}}%
%
\special{pn 20}%
\special{sh 1}%
\special{ar 1710 1880 10 10 0  6.28318530717959E+0000}%
\special{sh 1}%
\special{ar 1700 1880 10 10 0  6.28318530717959E+0000}%
%
\special{pn 20}%
\special{sh 1}%
\special{ar 3180 2380 10 10 0  6.28318530717959E+0000}%
\special{sh 1}%
\special{ar 3170 2380 10 10 0  6.28318530717959E+0000}%
\put(31.3000,-26.9000){\makebox(0,0)[lb]{$\frac{e_1}{2}$}}%
\put(14.2000,-20.4000){\makebox(0,0)[lb]{$\frac{e_2}{2}$}}%
\put(19.9000,-17.7000){\makebox(0,0)[lb]{$\frac{e_1+e_2}{2}$}}%
%
\special{pn 20}%
\special{sh 1}%
\special{ar 2010 2050 10 10 0  6.28318530717959E+0000}%
\special{sh 1}%
\special{ar 2000 2050 10 10 0  6.28318530717959E+0000}%
%
\special{pn 20}%
\special{sh 1}%
\special{ar 1390 1700 10 10 0  6.28318530717959E+0000}%
\special{sh 1}%
\special{ar 1380 1700 10 10 0  6.28318530717959E+0000}%
%
\special{pn 8}%
\special{pa 3760 2370}%
\special{pa 1996 1378}%
\special{dt 0.045}%
\end{picture}%
\end{center}
\end{figure}

$T^2$ is constructed by the $G_2$ lattice
whose basis vectors are $e_1 = 1$ and $e_2 = (-3+i\sqrt{3})/2$.
The orbifold $T^2/Z_6$ is obtained 
by further identifying $\varphi z$ with $z$ where $\varphi = e^{\pi i/3}$.
The resultant space is depicted in Figure \ref{F3}.
Basis vectors are transformed as $\varphi e_1=2e_1+e_2$, 
$\varphi e_2=-3e_1-e_2$ under the $Z_6$ transformation
$z \to \varphi z$. 
Fixed points are $z_{\text{fp}}=0$ for the $Z_6$ transformation
$z \to \varphi z$,
$z_{\text{fp}}=0$, $e_2/3$ and $2e_2/3$ for the $Z_3$ transformation 
$z \to \varphi^2 z$
and $z_{\text{fp}}=0$, $e_1/2$, $e_2/2$ and $(e_1+e_2)/2$ 
for the $Z_2$ transformation $z \to \varphi^3 z$,
and around these points we define ten kinds of transformations:
\begin{eqnarray}
&~& s_0: z\rightarrow \varphi z, ~~
s_{10}: z\rightarrow \varphi^2 z, ~~
s_{11}: z\rightarrow \varphi^2 z+e_1+e_2, ~~
s_{12}: z\rightarrow \varphi^2 z+2e_1+2e_2, \nonumber \\
&~& s_{20}: z\rightarrow \varphi^3 z, ~~
s_{21}: z\rightarrow \varphi^3 z+e_1, ~~
s_{22}: z\rightarrow \varphi^3 z+e_2, ~~
s_{23}: z\rightarrow \varphi^3 z+e_1+e_2, \nonumber \\
&~& t_1: z\rightarrow z+e_1, ~~
t_2: z\rightarrow z+e_2
\label{Z6-transf}
\end{eqnarray}
and they satisfy the relations;
\begin{eqnarray}
&~& s_0^6=s_{10}^3=s_{11}^3=s_{12}^3=s_{20}^2=s_{21}^2=s_{22}^2=s_{23}^2=I,
~s_{11}=t_1t_2s_{10},~s_{12}=t_1^2t_2^2s_{10},\nonumber \\
&~& s_{21}=t_1s_{20},~s_{22}=t_2s_{20},~s_{23}=t_1t_2s_{20}=s_{21}s_{20}s_{22}
=s_{22}s_{20}s_{21}=s_{11}s_0,\nonumber \\
&~& s_{10}=s_0^2,~s_{20}=s_0^3,~t_1t_2=t_2t_1,~
t_2 = s_0^2 t_1 s_0 t_1 s_0^3,\nonumber \\
&~& (s_0 s_{10})^4 = (s_0 s_{11})^4 = (s_0 s_{12})^4 = I,~
(s_0 s_{20})^3 = (s_0 s_{21})^3 = (s_0 s_{22})^3 = (s_0 s_{23})^3 = I.
\label{Z6-relations}
\end{eqnarray}
We denote the representation matrix for the $Z_6$ transformation
$s_0:z \to e^{\pi i/3} z$ as $U_0$ and other representation
matrices are determined uniquely, if $U_0$ is given.

\section{Fermions on six dimensions}
\label{app-B}

We explain gamma matrices, charge conjugation of fermions and
$Z_m$ transformation properties on six dimensions~\cite{K&M4}.
We use the metric $\eta_{MN} = \mbox{diag}(1, -1, -1, -1, -1, -1)$
 ($M, N=0, 1, 2, 3, 5, 6$),
and the following representation 
for six-dimensional gamma matrices:
\begin{eqnarray}
&~& \Gamma^{\mu} = \gamma^{\mu} \otimes \sigma^3
=\left(
\begin{array}{cc}
\gamma^{\mu} & 0 \\
0 & -\gamma^{\mu}
\end{array}
\right),~~
\Gamma^{5} = I_{4 \times 4}  \otimes i \sigma^1
=\left(
\begin{array}{cc}
0 & i I_{4 \times 4} \\
i I_{4 \times 4} & 0 
\end{array}
\right), 
\label{Gamma-5}\\
&~& \Gamma^{6} = I_{4 \times 4}  \otimes i \sigma^2
=\left(
\begin{array}{cc}
0 & I_{4 \times 4} \\
-I_{4 \times 4} & 0 
\end{array}
\right), 
\label{Gamma-6}
\end{eqnarray}
where $\mu=0, 1, 2, 3$,
$\sigma^i$ ($i=1, 2, 3$) are Pauli matrices,
and $I_{4 \times 4}$ is the $4 \times 4$ unit matrix.
We take the chiral representation on four-dimensional
space-time for $\gamma^{\mu}$ such that
\begin{eqnarray}
\gamma^{\mu}
\equiv 
\left(
\begin{array}{cc}
0 & \sigma^{\mu} \\
\overline{\sigma}^{\mu} & 0
\end{array}
\right),~~ \sigma^{\mu} = (I_{2 \times 2}, \sigma^{i}),~~
\overline{\sigma}^{\mu} = (I_{2 \times 2}, -\sigma^{i}),
\label{gamma-mu}
\end{eqnarray}
where $I_{2 \times 2}$ is the $2 \times 2$ unit matrix.
The $\Gamma^M$ satisfy the anti-commutation relations
of the Clifford algebra such that
$\{\Gamma^M, \Gamma^N\}=2 \eta^{MN}$
where $\eta^{MN}$ is the inverse of $\eta_{MN}$.
The chirality operator $\Gamma_7$ 
for six-dimensional fermion $\Psi$ is defined by
\begin{eqnarray}
\Gamma_7 \equiv \Gamma^0 \Gamma^1 \Gamma^2 \Gamma^3
 \Gamma^5 \Gamma^6 = -\gamma_{5} \otimes \sigma^3
=\left(
\begin{array}{cc}
-\gamma_5 & 0 \\
0 & \gamma_5
\end{array}
\right), 
\label{Gamma7}
\end{eqnarray}
where $\gamma_5$ is the chirality operator 
on four dimensions defined by
\begin{eqnarray}
\gamma_5 \equiv i \gamma^0 \gamma^1
\gamma^2 \gamma^3
\equiv 
\left(
\begin{array}{cc}
-I_{2 \times 2} & 0 \\
0 & I_{2 \times 2}
\end{array}
\right).
\label{gamma-5}
\end{eqnarray}
Six-dimensional fermions with a definite chirality
is called Weyl fermions on six dimensions.
The Weyl fermion $(\Psi_+)$ with positive chirality 
and that $(\Psi_-)$ with negative chirality
are given by
\begin{eqnarray}
&~& \Psi_+ = \frac{1+\Gamma_7}{2} \Psi 
= \left(
\begin{array}{cc}
\frac{1-\gamma_5}{2} & 0 \\
0 & \frac{1+\gamma_5}{2} 
\end{array}
\right)
\Psi
= \left( 
\begin{array}{c}
\psi_{+L} \\
\psi_{+R}
\end{array}
\right),
\label{Psi+}\\
&~& \Psi_- = \frac{1-\Gamma_7}{2} \Psi
= \left(
\begin{array}{cc}
\frac{1+\gamma_5}{2} & 0 \\
0 & \frac{1-\gamma_5}{2} 
\end{array}
\right)
\Psi
= \left( 
\begin{array}{c}
\psi_{-R} \\
\psi_{-L} 
\end{array}
\right),
\label{Psi-}
\end{eqnarray}
respectively.
Here, the subscript $\pm$ and $L(R)$
stand for the chiralities on six and four dimensions, respectively.
Using Weyl fermions $\xi_{\pm}$ and $\eta^*_{\pm}$
on four dimensions,
$\Psi$ and $\psi_{\pm L(R)}$ are expressed as
\begin{eqnarray}
\Psi = \left( 
\begin{array}{c}
\xi_+ \\
\eta^*_- \\
\xi_- \\
\eta^*_+
\end{array}
\right),~~
\psi_{+L} = \left( 
\begin{array}{c}
\xi_+ \\
0
\end{array}
\right),~~
\psi_{+R} = \left( 
\begin{array}{c}
0 \\
\eta^*_+
\end{array}
\right),~~
\psi_{-L} = \left( 
\begin{array}{c}
\xi_- \\
0
\end{array}
\right),~~
\psi_{-R} = \left( 
\begin{array}{c}
0 \\
\eta^*_-
\end{array}
\right).
\label{Psi-Weyls}
\end{eqnarray}

The charge conjugation of $\Psi$ is defined as
\begin{eqnarray}
\Psi^c \equiv B \Psi^*,
\label{CC-Psi}
\end{eqnarray}
where $B$ is a $8 \times 8$ matrix
which satisfies the relation
\begin{eqnarray}
B^{-1} \Gamma^{M} B = -\left(\Gamma^M\right)^*.
\label{B-relation}
\end{eqnarray}
The $B$ is given by
\begin{eqnarray}
B = -i \Gamma_7 \Gamma^2 \Gamma^5 
= \left(
\begin{array}{cccc}
0 & 0 & 0 & \sigma^2 \\
0 & 0 & \sigma^2 & 0 \\
0 & \sigma^2 & 0 & 0 \\
\sigma^2 & 0 & 0 & 0
\end{array}
\right)
\label{B}
\end{eqnarray}
up to a phase factor
and, using it, we derive the charge conjugation of 
$\xi_{\pm}$ and $\eta^*_{\pm}$,
\begin{eqnarray}
B\left( 
\begin{array}{c}
\xi_+ \\
0 \\
0 \\
0
\end{array}
\right)
=\left( 
\begin{array}{c}
0 \\
0 \\
0 \\
\sigma^2 \xi^*_+
\end{array}
\right),~~
B\left( 
\begin{array}{c}
0 \\
0 \\
0 \\
\eta^*_{+}
\end{array}
\right)
=\left( 
\begin{array}{c}
\sigma^2 \eta_+ \\
0 \\
0 \\
0
\end{array}
\right)
\label{Psi+CC}
\end{eqnarray}
and
\begin{eqnarray}
B\left( 
\begin{array}{c}
0 \\
0 \\
\xi_- \\
0
\end{array}
\right)
=\left( 
\begin{array}{c}
0 \\
\sigma^2 \xi^*_- \\
0 \\
0
\end{array}
\right),~~
B\left( 
\begin{array}{c}
0 \\
\eta^*_{-} \\
0 \\
0
\end{array}
\right)
=\left( 
\begin{array}{c}
0 \\
0 \\
\sigma^2 \eta_- \\
0
\end{array}
\right).
\label{Psi-CC}
\end{eqnarray}
From (\ref{Psi+CC}) and (\ref{Psi-CC}),
we find that the chirality in six dimensions
does not flip under the charge conjugation.

In terms of $\psi_{\pm L(R)}$,
the kinetic terms for $\Psi_+$ and $\Psi_-$ are rewritten as
\begin{eqnarray}
i \overline{\Psi}_+ \Gamma^M D_{M} \Psi_+
&=& i \overline{\Psi}_+ \Gamma^{\mu} D_{\mu} \Psi_+ 
+ i \overline{\Psi}_+ \Gamma^z D_{z} \Psi_+ 
+ i \overline{\Psi}_+ \Gamma^{\overline{z}} D_{\overline{z}} \Psi_+
\nonumber \\
&=& i \overline{\psi}_{+L} \gamma^{\mu} D_{\mu} \psi_{+L}
+ i \overline{\psi}_{+R} \gamma^{\mu} D_{\mu} \psi_{+R}
- 2\overline{\psi}_{+L} D_z \psi_{+R}
+ 2\overline{\psi}_{+R} D_{\overline{z}} \psi_{+L},
\label{Psi+kinetic}\\
i \overline{\Psi}_- \Gamma^M D_{M} \Psi_-
&=& i \overline{\Psi}_- \Gamma^{\mu} D_{\mu} \Psi_- 
+ i \overline{\Psi}_- \Gamma^z D_{z} \Psi_- 
+ i \overline{\Psi}_- \Gamma^{\overline{z}} D_{\overline{z}} \Psi_-
\nonumber \\
&=& i \overline{\psi}_{-R} \gamma^{\mu} D_{\mu} \psi_{-R}
+ i \overline{\psi}_{-L} \gamma^{\mu} D_{\mu} \psi_{-L}
- 2\overline{\psi}_{-R} D_z \psi_{-L}
+ 2\overline{\psi}_{-L} D_{\overline{z}} \psi_{-R},
\label{Psi-kinetic}
\end{eqnarray}
where $\overline{\Psi}_+$, $\overline{\Psi}_-$, $\Gamma^{z}$ and $\Gamma^{\overline{z}}$ are defined by
\begin{eqnarray}
\overline{\Psi}_+ &\equiv& \Psi_+^{\dagger} \Gamma^{0}
= \left(\psi^{\dagger}_{+L} \gamma^0, 
- \psi^{\dagger}_{+R} \gamma^0 \right) 
= \left(\overline{\psi}_{+L}, -\overline{\psi}_{+R} \right),
\nonumber \\
\overline{\Psi}_- &\equiv& \Psi_-^{\dagger} \Gamma^{0}
= \left(\psi^{\dagger}_{-R} \gamma^0, 
- \psi^{\dagger}_{-L} \gamma^0 \right) 
= \left(\overline{\psi}_{-R}, -\overline{\psi}_{-L} \right),
\label{6DoverlinePsi} \\ 
\Gamma^{z} &\equiv& 
\Gamma^{5} + i \Gamma^{6}
= 2i I_{4 \times 4}  \otimes \sigma_{+}
=\left(
\begin{array}{cc}
0 & 2i I_{4 \times 4} \\
0 & 0
\end{array}
\right),
\label{Gamma-z}\\
\Gamma^{\overline{z}} &\equiv& 
\Gamma^{5} - i \Gamma^{6}
= 2i I_{4 \times 4}  \otimes \sigma_{-}
=\left(
\begin{array}{cc}
0 & 0 \\
2i I_{4 \times 4} & 0
\end{array}
\right).
\label{Gamma-barz}
\end{eqnarray}
Here, $z \equiv x^5 + i x^6$ and $\overline{z} \equiv x^5 - i x^6$.
The Kaluza-Klein masses are generated 
from the terms including $D_z$ and $D_{\overline{z}}$
upon compactification.

The $Z_m$ elements are the eigenvalues of
the representation matrices $T_{\Psi_{\pm}}[U_a, \eta_{a\pm}]$
for the $Z_m$ transformation $z \to f_a(z)$
($\underbrace{f_a \circ f_a \cdot \circ f_a}_m(z) = z$),
operating $\Psi_{\pm}(x, z, \overline{z})$
such that
\begin{eqnarray}
\Psi_{\pm}(x, f_a(z), \overline{f_a}(\overline{z})) 
=T_{\Psi_{\pm}}[U_a, \eta_{a\pm}] \Psi_{\pm}(x, z, \overline{z}),
\label{TPhi}
\end{eqnarray}
where $U_a$ represent the representation matrices
for the fundamental representation,
$\eta_{a\pm}$ are the intrinsic $Z_m$ elements
and the subscript $L$ and $R$ are omitted on $\eta_{a\pm}$.
Let the intrinsic $Z_m$ elements of $\psi_{\pm L(R)}$
be $\eta_{a\pm L(R)}$.
Then, the intrinsic $Z_m$ elements of $\psi^{\dagger}_{\pm L(R)}$
are $\overline{\eta_{a\pm L(R)}}$
(complex conjugations of $\eta_{a\pm L(R)}$).
From the $Z_m$ invariance of the kinetic term (\ref{Psi+kinetic}) 
and (\ref{Psi-kinetic})
and the $Z_m$ transformation property of the covariant derivative 
$D_z \to \overline{\rho} D_z$ and 
$D_{\overline{z}} \to \rho D_{\overline{z}}$
under $z \to \rho z$
and $\overline{z} \to \overline{\rho}~\overline{z}$
($\rho = e^{2\pi i/m}$, $\overline{\rho} = e^{-2\pi i/m}$),
the following relations are derived: 
\begin{eqnarray}
\eta_{a+R} = {\rho} \eta_{a+L},~~
\eta_{a-R} = \overline{\rho} \eta_{a-L}.
\label{eta-psi}
\end{eqnarray}

\section{Flavor numbers and charge conjugation}
\label{app-C}

We give formulae for flavor numbers
from a fermion in $\overline{[N, k]}(=[N, N-k])$ 
and study the relationship
between flavor numbers
from a fermion in $\overline{[N, k]}$ 
and those from a fermion in $[N, k]$
from the viewpoint of charge conjugation.

Under the representation matrices $U_a$
with $p_1 = 3$ and $p_2 = 2$,
$[N, N-k]$ is decomposed as
\begin{eqnarray}
[N, N-k]
= \sum_{l_1 =0}^{N-k} \sum_{l_2 = 0}^{N-k-l_1} \sum_{l_3 = 0}^{N-k-l_1-l_2} 
\cdots \sum_{l_{n-1} = 0}^{N-k-l_1-\cdots -l_{n-2}}  
\left({{}_{3}C_{l_1}}, {{}_{2}C_{l_2}}, 
{{}_{p_3}C_{l_3}}, \cdots, {{}_{p_n}C_{l_n}}\right),
\label{NN-k(p1=3)}
\end{eqnarray}
where $\sum_{i=1}^n l_i = N-k$.
From $[N, N-k] = \overline{[N, k]}$, 
hereafter we use the decomposition of $\overline{[N, k]}$ such that
\begin{eqnarray}
\overline{[N, k]}
= \sum_{l_1 =0}^{k} \sum_{l_2 = 0}^{k-l_1} \sum_{l_3 = 0}^{k-l_1-l_2} 
\cdots \sum_{l_{n-1} = 0}^{k-l_1-\cdots -l_{n-2}}  
\left(\overline{{}_{3}C_{l_1}}, \overline{{}_{2}C_{l_2}}, 
\overline{{}_{p_3}C_{l_3}}, \cdots, \overline{{}_{p_n}C_{l_n}}\right),
\label{overlineNk(p1=3)}
\end{eqnarray}
where $\sum_{i=1}^n l_i = k$.
Using the survival hypothesis and 
the equivalence on charge conjugation in four dimensions, 
we define the flavor number of each chiral fermion as
\begin{eqnarray}
&~& n_{\bar{d}} \equiv 
\left(\sharp (\overline{{}_{3}C_{2}}, \overline{{}_{2}C_{2}})_R 
- \sharp  (\overline{{}_{3}C_{1}}, \overline{{}_{2}C_{0}})_R\right) 
- \left(\sharp (\overline{{}_{3}C_{2}}, \overline{{}_{2}C_{2}})_L
- \sharp  (\overline{{}_{3}C_{1}}, \overline{{}_{2}C_{0}})_L\right),  
\label{nd-bar}\\
&~& n_{l} \equiv 
\left(\sharp  (\overline{{}_{3}C_{3}}, \overline{{}_{2}C_{1}})_R 
- \sharp  (\overline{{}_{3}C_{0}}, \overline{{}_{2}C_{1}})_R\right)
- \left(\sharp  (\overline{{}_{3}C_{3}}, \overline{{}_{2}C_{1}})_L 
- \sharp  (\overline{{}_{3}C_{0}}, \overline{{}_{2}C_{1}})_L\right),  
\label{nl-bar}\\
&~& n_{\bar{u}} \equiv 
\left(\sharp  (\overline{{}_{3}C_{2}}, \overline{{}_{2}C_{0}})_R  
- \sharp  (\overline{{}_{3}C_{1}}, \overline{{}_{2}C_{2}})_R\right)
- \left(\sharp  (\overline{{}_{3}C_{2}}, \overline{{}_{2}C_{0}})_L
- \sharp  (\overline{{}_{3}C_{1}}, \overline{{}_{2}C_{2}})_L\right),  
\label{nu-bar}\\
&~& n_{\bar{e}} \equiv 
\left(\sharp  (\overline{{}_{3}C_{0}}, \overline{{}_{2}C_{2}})_R  
- \sharp  (\overline{{}_{3}C_{3}}, \overline{{}_{2}C_{0}})_R\right) 
- \left(\sharp  (\overline{{}_{3}C_{0}}, \overline{{}_{2}C_{2}})_L 
- \sharp  (\overline{{}_{3}C_{3}}, \overline{{}_{2}C_{0}})_L\right),  
\label{ne-bar}\\
&~& n_{q} \equiv 
\left(\sharp  (\overline{{}_{3}C_{1}}, \overline{{}_{2}C_{1}})_R  
- \sharp  (\overline{{}_{3}C_{2}}, \overline{{}_{2}C_{1}})_R\right) 
- \left(\sharp  (\overline{{}_{3}C_{1}}, \overline{{}_{2}C_{1}})_L
- \sharp  (\overline{{}_{3}C_{2}}, \overline{{}_{2}C_{1}})_L\right),  
\label{nq-bar}
\end{eqnarray}
where $\sharp$ represents the number of zero modes
for each multiplet.
The total number of neutrino singlets $(\nu_{R})^c$ 
and/or $\nu_R$ is defined as
\begin{eqnarray}
n_{\bar{\nu}} \equiv \sharp  (\overline{{}_{3}C_{0}}, \overline{{}_{2}C_{0}})_R  
+ \sharp  (\overline{{}_{3}C_{3}}, \overline{{}_{2}C_{2}})_R 
+ \sharp  (\overline{{}_{3}C_{0}}, \overline{{}_{2}C_{0}})_L 
+ \sharp  (\overline{{}_{3}C_{3}}, \overline{{}_{2}C_{2}})_L.  
\label{nnu-bar}
\end{eqnarray}
Note that we have relations:
\begin{eqnarray}
(\overline{{}_{3}C_{l_1}}, \overline{{}_{2}C_{l_2}})_{R(L)}
= ({}_{3}C_{3-l_1}, {}_{2}C_{2-l_2})_{R(L)}.  
\label{barC=C}
\end{eqnarray}

Formulae for the SM species and neutrino singlets
derived from a pair of six-dimensional Weyl fermions
$(\Psi_{+}, \Psi_{-})$ in $\overline{[N, k]}$
are given by
\begin{eqnarray}
\left. n_{\bar{d}}\right|_{\overline{[N, k]}} 
&=& \sum_{\pm} \sum_{(l_1, l_2) = (2,2),(1,0)} 
\sum_{l_3 = 0}^{k-l_1-l_2} 
\cdots \sum_{l_{n-1} = 0}^{k-l_1-\cdots -l_{n-2}} 
(-1)^{l_1+l_2} \tilde{P}_{mk\pm}~
{}_{p_3}C_{l_3} \cdots {}_{p_{n}}C_{l_{n}}, 
\label{nd-ZM-bar}\\
\left. n_{l}\right|_{\overline{[N, k]}} &=& 
\sum_{\pm} \sum_{(l_1, l_2) = (3,1),(0,1)} 
\sum_{l_3 = 0}^{k-l_1-l_2} 
\cdots \sum_{l_{n-1} = 0}^{k-l_1-\cdots -l_{n-2}} 
(-1)^{l_1+l_2} \tilde{P}_{mk\pm}~
{}_{p_3}C_{l_3} \cdots {}_{p_{n}}C_{l_{n}}, 
\label{nl-ZM-bar}\\
\left. n_{\bar{u}}\right|_{\overline{[N, k]}} 
&=& \sum_{\pm} \sum_{(l_1, l_2) = (2,0),(1,2)} 
\sum_{l_3 = 0}^{k-l_1-l_2} 
\cdots \sum_{l_{n-1} = 0}^{k-l_1-\cdots -l_{n-2}} 
(-1)^{l_1+l_2} \tilde{P}_{mk\pm}~
{}_{p_3}C_{l_3} \cdots {}_{p_{n}}C_{l_{n}}, 
\label{nu-ZM-bar}\\
\left. n_{\bar{e}}\right|_{\overline{[N, k]}} 
&=& \sum_{\pm} \sum_{(l_1, l_2) = (0,2),(3,0)} 
\sum_{l_3 = 0}^{k-l_1-l_2} 
\cdots \sum_{l_{n-1} = 0}^{k-l_1-\cdots -l_{n-2}} 
(-1)^{l_1+l_2} \tilde{P}_{mk\pm}~
{}_{p_3}C_{l_3} \cdots {}_{p_{n}}C_{l_{n}}, 
\label{ne-ZM-bar}\\
\left. n_{q}\right|_{\overline{[N, k]}} 
&=& \sum_{\pm} \sum_{(l_1, l_2) = (1,1),(2,1)} 
\sum_{l_3 = 0}^{k-l_1-l_2} 
\cdots \sum_{l_{n-1} = 0}^{k-l_1-\cdots -l_{n-2}} 
(-1)^{l_1+l_2} \tilde{P}_{mk\pm}~
{}_{p_3}C_{l_3} \cdots {}_{p_{n}}C_{l_{n}}, 
\label{nq-ZM-bar}\\
\left. n_{\bar{\nu}}\right|_{\overline{[N, k]}} 
&=& \sum_{\pm} \sum_{(l_1, l_2) = (0,0),(3,2)}
\sum_{l_3 = 0}^{k-l_1-l_2} 
\cdots \sum_{l_{n-1} = 0}^{k-l_1-\cdots -l_{n-2}} 
\tilde{P}_{mk}^{(\nu)}~{}_{p_3}C_{l_3} \cdots {}_{p_{n}}C_{l_{n}}, 
\label{nnu-ZM-bar}
\end{eqnarray}
where $\tilde{P}_{mk\pm}$ and $\tilde{P}_{mk\pm}^{(\nu)}$ are defined by
\begin{eqnarray}
\tilde{P}_{mk\pm} \equiv \tilde{P}_{mk\pm R} - \tilde{P}_{mk\pm L},~~
\tilde{P}_{mk\pm}^{(\nu)} \equiv \tilde{P}_{mk\pm R} + \tilde{P}_{mk\pm L},
\label{tildePMk}
\end{eqnarray}
respectively.
The $\tilde{P}_{mk\pm R(L)}$ are projection operators
to pick out zero modes of $\psi_{\pm R(L)}$ in $\overline{[N, k]}$,
and they are listed in Table \ref{projection-tilde}.
\begingroup
\renewcommand{\arraystretch}{1.2}
\begin{table}[htb]
\caption{The projection operators $\tilde{P}_{mk\pm R(L)}$.}
\label{projection-tilde}
\begin{center}
\begin{tabular}{c|c|c|c|c} \hline
$T^2/Z_m$ & $\tilde{P}_{mk+R}$ & $\tilde{P}_{mk+L}$ 
& $\tilde{P}_{mk-R}$ & $\tilde{P}_{mk-L}$
\\ \hline\hline
$T^2/Z_2$ & $\tilde{P}_{2k+}^{(1,1,1)}$ & $\tilde{P}_{2k+}^{(-1,-1,-1)}$
& $\tilde{P}_{2k-}^{(1,1,1)}$ & $\tilde{P}_{2k-}^{(-1,-1,-1)}$
\\ \hline
$T^2/Z_3$ & $\tilde{P}_{3k+}^{(1,1)}$ 
& $\tilde{P}_{3k+}^{(\omega, \omega)}$
& $\tilde{P}_{3k-}^{(1,1)}$
& $\tilde{P}_{3k-}^{(\overline{\omega}, \overline{\omega})}$
\\ \hline
$T^2/Z_4$ & $\tilde{P}_{4k+}^{(1, 1)}$
& $\tilde{P}_{4k+}^{(i,i)}$
& $\tilde{P}_{4k-}^{(1, 1)}$ & $\tilde{P}_{4k-}^{(-i, -i)}$ 
\\ \hline
$T^2/Z_6$ & $\tilde{P}_{6k+}^{(1)}$
& $\tilde{P}_{6k+}^{(\varphi)}$ 
& $\tilde{P}_{6k-}^{(1)}$
& $\tilde{P}_{6k-}^{(\overline{\varphi})}$ 
\\ \hline
\end{tabular}
\end{center}
\end{table}
\endgroup
In Table \ref{projection-tilde}, each operator is defined by
\begin{eqnarray}
&~& \tilde{P}_{2k\pm}^{\left((-1)^{n_0}, (-1)^{n_1}, (-1)^{n_2}\right)} \equiv 
\frac{1}{8}\left\{1 + (-1)^{n_0} \tilde{\mathcal{P}}^{(k)}_{0 \pm}\right\}
\left\{1 + (-1)^{n_1} \tilde{\mathcal{P}}^{(k)}_{1 \pm}\right\}
\left\{1 + (-1)^{n_2} \tilde{\mathcal{P}}^{(k)}_{2 \pm}\right\},
\label{P2-tilde}\\
&~& \tilde{P}_{3k\pm}^{\left(\omega^{n_0}, \omega^{n_1}\right)} 
\equiv \frac{1}{9}
\left\{1+ \overline{\omega}^{n_0} \tilde{\mathcal{P}}^{(k)}_{0 \pm} 
+ \overline{\omega}^{2n_0} 
\left(\tilde{\mathcal{P}}^{(k)}_{0 \pm}\right)^2\right\}
\left\{1+ \overline{\omega}^{n_1} 
\tilde{\mathcal{P}}^{(k)}_{1 \pm} 
+ \overline{\omega}^{2n_1} \left(
\tilde{\mathcal{P}}^{(k)}_{1 \pm}\right)^2\right\},
\label{P3-tilde}\\
&~& \tilde{P}_{4k\pm}^{\left(i^{n_0}, (-1)^{n_1}\right)} 
\equiv \frac{1}{16}
\left\{1+ (-i)^{n_0} \tilde{\mathcal{P}}^{(k)}_{0 \pm} 
+ (-i)^{2n_0} \left(\tilde{\mathcal{P}}^{(k)}_{0 \pm}\right)^2
+ (-i)^{3n_0} \left(\tilde{\mathcal{P}}^{(k)}_{0 \pm}\right)^3\right\}
\nonumber \\
&~& ~~~~~~~~~~~~~~~~~~~~~~~~~~~~~~~ \times
\left\{1+ (-i)^{n_1} \tilde{\mathcal{P}}^{(k)}_{1 \pm} 
+ (-i)^{2n_1} \left(\tilde{\mathcal{P}}^{(k)}_{1 \pm}\right)^2
+ (-i)^{3n_1} \left(\tilde{\mathcal{P}}^{(k)}_{1 \pm}\right)^3\right\},
\label{P4-tilde}\\
&~& \tilde{P}_{6k\pm}^{\left(\varphi^{n_0}\right)} 
\equiv \frac{1}{6}
\left\{1+ \overline{\varphi}^{n_0} \tilde{\mathcal{P}}^{(k)}_{0 \pm} 
+ \overline{\varphi}^{2n_0} \left(\tilde{\mathcal{P}}^{(k)}_{0 \pm}\right)^2
+ \overline{\varphi}^{3n_0} \left(\tilde{\mathcal{P}}^{(k)}_{0 \pm}\right)^3
+ \overline{\varphi}^{4n_0} \left(\tilde{\mathcal{P}}^{(k)}_{0 \pm}\right)^4
+ \overline{\varphi}^{5n_0} \left(\tilde{\mathcal{P}}^{(k)}_{0 \pm}\right)^5
\right\},
\label{P6-tilde}
\end{eqnarray}
where $\tilde{\mathcal{P}}^{(k)}_{a \pm}$ are the $Z_m$ elements
For instance, $\tilde{P}_{3k\pm}^{\left(\omega^{n_0}, \omega^{n_1}\right)}$
is an projection operator to pick out
modes with $\tilde{\mathcal{P}}^{(k)}_{0 \pm} = \omega^{n_0}$
and $\tilde{\mathcal{P}}^{(k)}_{1 \pm} = \omega^{n_1}$ in $\Psi_{\pm}$.
By the insertion of $(-1)^{l_1+l_2}$, 
we obtain
$\sharp (\overline{{}_{3}C_{l_1}}, \overline{{}_{2}C_{l_2}})_{R(L)}$
for $l_1 + l_2=$ even integer
and $-\sharp (\overline{{}_{3}C_{l_1}}, \overline{{}_{2}C_{l_2}})_{R(L)}$
for $l_1 + l_2=$ odd integer.

The $\tilde{\mathcal{P}}^{(k)}_{a \pm}$ of
$\left(\overline{{}_{3}C_{l_1}}, 
\overline{{}_{2}C_{l_2}}, \cdots, \overline{{}_{p_n}C_{l_n}}\right)$
are given by
\begin{eqnarray}
\tilde{\mathcal{P}}^{(k)}_{0 \pm}
= (-1)^{l_1+l_2+l_3+l_4 -k} \tilde{\eta}^{(k)}_{0 \pm},~~ 
\tilde{\mathcal{P}}^{(k)}_{1 \pm} 
= (-1)^{l_1+l_2+l_5+l_6-k} \tilde{\eta}^{(k)}_{1 \pm},~~
\tilde{\mathcal{P}}^{(k)}_{2 \pm}  
= (-1)^{l_1+l_3+l_5+l_7-k} \tilde{\eta}^{(k)}_{2 \pm}
\label{Z2-ele-tilde}
\end{eqnarray}
for (\ref{Z2-U}),
\begin{eqnarray}
\tilde{\mathcal{P}}^{(k)}_{0 \pm}
= \overline{\omega}^{l_1+l_2+l_3+2(l_4 + l_5 + l_6) -k}
\tilde{\eta}^{(k)}_{0 \pm},~~ 
\tilde{\mathcal{P}}^{(k)}_{1 \pm} 
= \overline{\omega}^{l_1+l_4+l_7+2(l_2 + l_5 + l_8) -k}
\tilde{\eta}^{(k)}_{1 \pm}
\label{Z3-ele-tilde}
\end{eqnarray}
for (\ref{Z3-U}),
\begin{eqnarray}
\tilde{\mathcal{P}}^{(k)}_{0 \pm}
= (-i)^{l_1+l_2+2(l_3+l_4) + 3(l_5 + l_6) -k}
\tilde{\eta}^{(k)}_{0 \pm},~~ 
\tilde{\mathcal{P}}^{(k)}_{1 \pm} 
=  (-i)^{l_1+l_6+2(l_4+l_7) + 3(l_2 + l_5) -k}
\tilde{\eta}^{(k)}_{1 \pm}
\label{Z4-ele-tilde}
\end{eqnarray}
for (\ref{Z4-U}) and
\begin{eqnarray}
\tilde{\mathcal{P}}^{(k)}_{0 \pm}
= \overline{\varphi}^{l_1+2 l_2+ 3 l_3+ 4 l_4 + 5 l_5
 -k} \tilde{\eta}^{(k)}_{0 \pm}
\label{Z6-ele-tilde}
\end{eqnarray}
for (\ref{Z6-U}).
The subscripts $L$ and $R$ 
on the intrinsic $Z_m$ elements are omitted
in (\ref{Z2-ele-tilde}),  (\ref{Z3-ele-tilde}), 
(\ref{Z4-ele-tilde}) and (\ref{Z6-ele-tilde}).
Notice that complex values $\omega$, $i$ and $\varphi$
in (\ref{Z3-ele}), (\ref{Z4-ele}) and (\ref{Z6-ele})
are replaced into their complex conjugated ones
in (\ref{Z3-ele-tilde}), (\ref{Z4-ele-tilde}) and (\ref{Z6-ele-tilde}),
because $U_a^{*}$ operate fields multiple times in place of $U_a$.

From (\ref{rhopsi}),
$\tilde{\eta}^{(k)}_{a \pm L}$ are determined from 
$\tilde{\eta}^{(k)}_{a \pm R}$ as
\begin{eqnarray}
\tilde{\eta}^{(k)}_{a + L} = \overline{\rho} \tilde{\eta}^{(k)}_{a + R},~~
\tilde{\eta}^{(k)}_{a - L} = {\rho} \tilde{\eta}^{(k)}_{a - R}.
\label{etaL-tilde}
\end{eqnarray}

In case that $\tilde{\eta}^{(k)}_{a \pm R} = 
\overline{{\eta}^{(k)}_{a \pm L}}$, 
we have the relations:
\begin{eqnarray}
\tilde{\mathcal{P}}^{(k)}_{a \pm}
= \overline{\mathcal{P}^{(k)}_{a \pm}}
\label{tilderho=barrho}
\end{eqnarray}
and derive the relations:
\begin{eqnarray}
&~& \tilde{P}_{2k\pm}^{\left((-1)^{n_0}, (-1)^{n_1}, (-1)^{n_2}\right)}
= {P}_{2k\pm}^{\left((-1)^{n_0}, (-1)^{n_1}, (-1)^{n_2}\right)},~~
\tilde{P}_{mk\pm}^{\left(\rho^{n_0}, \rho^{n_1}\right)}
= \overline{{P}_{mk\pm}^{\left(\overline{\rho}^{n_0}, 
\overline{\rho}^{n_1}\right)}}
= {P}_{mk\pm}^{\left(\overline{\rho}^{n_0}, \overline{\rho}^{n_1}\right)}
~~~(m=3, 4),
\nonumber \\
&~& \tilde{P}_{6k\pm}^{\left(\varphi^{n_0}\right)}
= \overline{{P}_{6k\pm}^{\left(\overline{\varphi}^{n_0}\right)}}
= {P}_{6k\pm}^{\left(\overline{\varphi}^{n_0}\right)}.
\label{tildeP=barP}
\end{eqnarray}
In the last equality in the above second relation,
we use the fact that the projection operators take a real number
$1$ or $0$.
From (\ref{tildeP=barP}), we find
that the flavor numbers derived from the projection by 
$(-1)^{l_1+l_2} \tilde{P}_{mk\pm}$
are equal to those from that by
$(-1)^{l_1+l_2} \overline{P_{mk\pm}} = (-1)^{l_1+l_2} P_{mk\pm}$.
In this way, we have a feature
that {\it each flavor number from a fermion in $[N,k]$ 
with intrinsic $Z_m$ elements $\eta^{(k)}_{a \pm}$
is equal to that from a fermion in $\overline{[N, k]}(=[N,N-k])$ 
with those satisfying $\tilde{\eta}^{(k)}_{a \pm R} = 
\overline{{\eta}^{(k)}_{a \pm L}}$ $($appropriate 
$\eta^{(N-k)}_{a \pm})$.}
In other words, {\it there is a one-to-one 
correspondence between
zero modes from a Weyl fermion in $[N,k]$ with $\eta^{(k)}_{a \pm}$
and those from a Weyl fermion in $[N,N-k]$ with
appropriate $\eta^{(N-k)}_{a \pm}$.}

Finally, let us obtain appropriate $\eta^{(N-k)}_{a \pm}$
to hold the above-stated correspondence, 
in the case with (\ref{Z3-U}) of $T^2/Z_3$.
In this case, $\mathcal{P}^{(N-k)}_{a \pm}$ are given by
\begin{eqnarray}
\mathcal{P}^{(N-k)}_{0 \pm}
= \omega^{l_1+l_2+l_3+2(l_4 + l_5 + l_6) -(N-k)}\eta^{(N-k)}_{0 \pm},~~ 
\mathcal{P}^{(N-k)}_{1 \pm} 
= \omega^{l_1+l_4+l_7+2(l_2 + l_5 + l_8) -(N-k)}\eta^{(N-k)}_{1 \pm}.
\label{Z3-ele-N-k}
\end{eqnarray}
By replacing $l_i$ into $p_i - l_i$ in $\mathcal{P}^{(N-k)}_{0 \pm}$
and $\mathcal{P}^{(N-k)}_{1 \pm}$,
we obtain $\tilde{\mathcal{P}}^{(k)}_{0 \pm}$
and $\tilde{\mathcal{P}}^{(k)}_{1 \pm}$ such that
\begin{eqnarray}
&~& \tilde{\mathcal{P}}^{(k)}_{0 \pm}
= \overline{\omega}^{l_1+l_2+l_3+2(l_4 + l_5 + l_6) -k}
\omega^{p_1 + p_2 + p_3 + 2(p_4 + p_5 + p_6) -N}
{\eta}^{(N-k)}_{0 \pm},~~
\label{Z3-ele-tilde-0} \\
&~& \tilde{\mathcal{P}}^{(k)}_{1 \pm} 
= \overline{\omega}^{l_1+l_4+l_7+2(l_2 + l_5 + l_8) -k}
\omega^{p_1 + p_4 + p_7 + 2(p_2 + p_5 + p_8) -N}
{\eta}^{(N-k)}_{1 \pm}.
\label{Z3-ele-tilde-1}
\end{eqnarray} 
Using (\ref{Z3-ele-tilde}), (\ref{Z3-ele-tilde-0}), (\ref{Z3-ele-tilde-1})
and $\tilde{\eta}^{(k)}_{a \pm R} = \overline{{\eta}^{(k)}_{a \pm L}}$,
we derive the relations:
\begin{eqnarray}
&~& \tilde{\eta}^{(k)}_{0 \pm R}
= \omega^{p_1 + p_2 + p_3 + 2(p_4 + p_5 + p_6) -N}
{\eta}^{(N-k)}_{0 \pm R} = \overline{\eta^{(k)}_{0 \pm L}},
\label{Z3-eta-tilde-0} \\
&~& \tilde{\eta}^{(k)}_{1 \pm R} 
= \omega^{p_1 + p_4 + p_7 + 2(p_2 + p_5 + p_8) -N}
{\eta}^{(N-k)}_{1 \pm R} = \overline{\eta^{(k)}_{1 \pm L}}.
\label{Z3-eta-tilde-1}
\end{eqnarray}
The equivalence based on
the relations (\ref{Z3-eta-tilde-0}) and (\ref{Z3-eta-tilde-1})
is illustrated with the particle contents listed in Table \ref{Table:Z3-1} 
and \ref{Table:Z3-2}.

\end{document}